\documentclass[
nofootinbib,
superscriptaddress,
 amsmath,amssymb,
 aps,
 floats,
 twocolumn,
prb,
floatfix,
]{revtex4-2}

\usepackage{mathtools}
\usepackage{xcolor}
\usepackage{graphicx}
\usepackage{dcolumn}
\usepackage{bm}
\usepackage{bbm}
\usepackage{physics}%
\usepackage{booktabs}
\definecolor{TUblue}{RGB}{0, 102, 153}
\definecolor{TUred}{RGB}{186, 70, 130}
\usepackage[pdfencoding=auto, 
            psdextra,
            colorlinks=true,
            citecolor={TUred},
            linkcolor={TUblue}
            ]{hyperref}
\usepackage{orcidlink}
\usepackage{textgreek}
\usepackage[capitalise]{cleveref}
\usepackage[normalem]{ulem} 

\usepackage{tikz}
\usepackage{tikz-feynhand}
\usepackage[normalem]{ulem}
\usepackage{floatrow}
\usepackage{placeins} 
\usepackage[caption=false]{subfig}
\newcommand{\ii}{\mathrm{i}}
\newcommand{\e}{\mathrm{e}}
\newcommand{\mv}{\mathrm{v}}

\newcommand{\pdg}{{\ensuremath{\phantom{\dagger}}}}

\newcommand{\up}{\uparrow}
\newcommand{\down}{\downarrow}
\newcommand{\vk}{\ensuremath{\mathbf{k}}}
\newcommand{\baralpha}{\ensuremath{\overline{\alpha}}}
\newcommand{\vq}{\ensuremath{\mathbf{q}}}
\newcommand{\fq}{\ensuremath{\mathbf{\mathfrak{q}}}}

\newcommand{\ph}{\ensuremath{{ph}}}

\newcommand{\bchi}{\ensuremath{\bm{\chi}}}
\newcommand{\bv}{\ensuremath{\bm{\tilde{v}}}}
\renewcommand{\Tr}{\operatorname{Tr}}
\renewcommand{\Re}{\operatorname{Re}}
\renewcommand{\Im}{\operatorname{Im}}

\newcommand{\create}[1]{\hat{c}_{#1}^\dagger}
\newcommand{\ancreate}[1]{\hat{c}_{#1}^{(\dagger)}}
\newcommand{\annihil}[1]{\hat{c}_{#1}^{\pdg}}

\definecolor{colorA}{RGB}{89, 110, 127}
\definecolor{colorB}{RGB}{232, 178, 143}

\newcommand\varmp{\mathbin{\vcenter{\hbox{%
  \oalign{$\scriptscriptstyle ({+})$\cr
          \noalign{\kern-.3ex}
          \hfil$\scriptscriptstyle -$\hfil\cr}%
}}}}
\newcommand{\stackeq}[1]{\stackrel{\mathclap{\scriptscriptstyle #1}}{=}}
\DeclareRobustCommand\fermion{
    \begin{tikzpicture}[]
        \begin{feynhand}
         \vertex (a) at (-0.4,0);
         \vertex (b) at (0.4,0);
        \propag [fermion] (a) to (b);
        \end{feynhand}
        \end{tikzpicture}
}
\DeclareRobustCommand\vertex{
\begin{tikzpicture}
\begin{feynhand}
    \draw (-0.1,-0.1) to (0,0){};
    \draw (0.1,-0.1) to (0,0){};
    \draw (-0.1,0.1) to (0,0){};
    \draw (0.1,0.1) to (0,0){};
    \vertex [gray,dot] at (0,0){};
\end{feynhand}
\end{tikzpicture}
}

\begin{document}

\title{Non-Perturbative Feats in the Physics of Correlated Antiferromagnets}
\author{M.~Reitner\,\orcidlink{0000-0002-2529-0847}}\email{matthias.reitner@outlook.com}
\affiliation{Institute of Solid State Physics, TU Wien, 1040 Vienna, Austria}
\author{L.~Del~Re\,\orcidlink{0000-0002-8765-2866}}
\affiliation{Institut f\"ur Theoretische Physik und Astrophysik and W\"urzburg-Dresden Cluster
of Excellence ct.qmat, Universit\"at W\"urzburg, 97074 W\"urzburg, Germany}
\affiliation{Max-Planck-Institut f\"ur Festk\"orperforschung, Heisenbergstra{\ss}e 1, 70569 Stuttgart, Germany}
\author{M.~Capone\,\orcidlink{0000-0002-9811-5089}}
\affiliation{International School for Advanced Studies (SISSA), via Bonomea 265, 34136 Trieste, Italy}
\affiliation{CNR-IOM, Istituto Officina dei Materiali, Consiglio Nazionale delle Ricerche, Via Bonomea 265, 34136 Trieste, Italy}
\author{A.~Toschi\,\orcidlink{0000-0001-5669-3377}}
\affiliation{Institute of Solid State Physics, TU Wien, 1040 Vienna, Austria}

\date{\today}


\begin{abstract}
In the last decades multifaceted manifestations of the breakdown of the self-consistent perturbation theory have been identified for the many-electron problem. Yet, the investigations have been so far mostly limited to paramagnetic states, where symmetry breaking is not allowed.
Here, we extend the analysis to the spontaneously symmetry-broken antiferromagnetic (AF) phase of the repulsive Hubbard model. 
To this aim, we calculated two-particle quantities using dynamical mean-field theory for the AF-ordered Hubbard model and studied the possible occurrence of divergences of the irreducible vertex functions in the charge and spin sectors.  
Our calculations pinpoint the divergences in the AF phase diagram, showing that while the onset of AF order mitigates the breakdown of the perturbation expansion, it does not fully prevent it.
Moreover, we have been able to link the changes in the dynamical structure of the corresponding generalized susceptibilities to the physical crossover from a weak-coupling (Slater) to a strong-coupling (Heisenberg) antiferromagnet, which takes place as the interaction strength is gradually increased. 
Finally, we discuss possible physical consequences of the irreducible vertex divergences in triggering phase-separation instabilities within the AF phase and elaborate on the implications of our findings for two-dimensional systems, where the onset of a long-range AF order is prevented by the Mermin-Wagner theorem.
\end{abstract}

\maketitle

\vskip 5mm

\section{Introduction}

Several of the most fascinating phenomena in condensed matter theory, such as unconventional high-temperature superconductivity, quantum criticality \cite{loehneysen2007,schafer2017}, or the appearance of non-trivial features in spectroscopic experiments (pseudogap \cite{timusk1999,gunnarsson2015,worm2024}, waterfalls \cite{krsnik2025}, hour-glasses, to cite a few) occur in materials in which the screening of the on-site electron-electron interaction works rather poorly. As a result, the quantum field theoretical description of the underlying many-electron physics needs to be pursued in parameter regimes where the validity of the many-body perturbation expansion is not guaranteed a priori.
This arguably explains the efforts made in the last decade by several research groups to investigate all multi-faceted manifestations of the breakdown of the self-consistent perturbation theory, both on a formal \cite{schafer2013,kozik2015,stan2015,gunnarsson2016,gunnarsson2017,chalupa2018,tarantino2018,vucicevic2018,springer2020,reitner2024,essl2024}  and a physical \cite{reitner2020,chalupa2021,mazitov2022localmomentA,mazitov2022localmomentB,adler2024,kowalski2024} level.

In a nutshell, we recall that the major manifestations of such breakdown are the misleading convergence of the self-consistent (bold) perturbation expansion \cite{kozik2015,stan2015,tarantino2018,vucicevic2018} in broad parameter regimes of the many-electron problem, as well as the associated \cite{gunnarsson2017} divergences of the irreducible two-particle vertex functions \cite{schafer2013,schafer2016,gunnarsson2016,springer2020,pelz2023,essl2024}. While the relevance of these non-perturbative features for algorithmic schemes based on resummations of bold Feynman diagrams or on parquet equations is immediately evident, recently a clear link between such formal aspects of the many-electron theory and the underlying physical properties has been identified. In particular, the breakdown of the self-consistent perturbation expansion appears to be directly driven \cite{chalupa2021,mazitov2022localmomentA,mazitov2022localmomentB,adler2024} by one of the major hallmarks of the correlated electron physics, i.e.~the gradual formation of local magnetic moments and the {\sl simultaneous} suppression of on-site charge fluctuations. Furthermore, under certain conditions, its occurrence can even reverse the sign of the effective electronic interaction at low energies, turning it into an attraction \cite{nourafkan2019,reitner2020}. The latter  
has been shown to trigger phase-separation instabilities in (purely repulsive) electronic systems~\cite{reitner2020,reitner2024,kowalski2024}, and, prospectively, it can lead to significant renormalization of the electron-phonon coupling~\cite{Grilli1994,Capone2010} and/or of the electronic pairing~\cite{Grilli1991PRL,Emery1993,Dagotto1994} in the presence of strong-correlations. 

The investigation of these issues has been typically carried out, non-perturbatively, by means of analytical calculations \cite{kozik2015,stan2015,schafer2016,gunnarsson2017,tarantino2018,essl2024} of simplified models, of numerically precise treatments \cite{gunnarsson2016,chalupa2018,chalupa2021,rohshap2024} of single-impurity Anderson models and, above all, by solving, on the two-particle level \cite{schafer2013,kozik2015,schafer2016,gunnarsson2016,vucicevic2018,springer2020,reitner2020,chalupa2021,pelz2023,adler2024,kowalski2024}, the Hubbard model via dynamical mean-field theory (DMFT) \cite{georges1996} or its cluster extensions \cite{maier2005}. 

Hitherto, however, all these studies \footnote{We note here the partial exception of Ref.~\cite{essl2024}, where the SU(2)-symmetry is explicitly broken by an external magnetic field in the basic case of an Hubbard atom.} were performed in the paramagnetic phase, i.e., not allowing for any spontaneous breaking of the SU(2)-symmetry {\sl even} in situations where the low-temperature phase of the system considered is known to display a thermodynamically stable antiferromagnetic (AF) long-range ordering.

Indeed, despite the higher numerical effort of performing two-particle calculations in a lower-symmetry phase, an extension of the previous investigations to the AF ordered regime would be important for several reasons. 

On a fundamental level, the study of the breakdown of the bold perturbation expansion in the presence of a (thermodynamically stable) AF ordering could allow for a precise understanding of the competitive interplay between long-range order and electronic  correlation, beyond the intriguing effects already observed at the one-particle level \cite{sangiovanni2006,taranto2012} that highlighted the crucial relevance of many-body effects beyond mean-field in the AF phase.

On a more physical level, such a study could unveil whether the onset of long-range order can mitigate the breakdown of perturbation theory, following the intuition that the singularities of the theory can be weakened in the broken symmetry AF state. Further, our analysis could clarify the underlying relations between the breakdown of the perturbation theory and the different regimes of a correlated antiferromagnet, which typically evolves from a Slater AF at weak-coupling (whose stabilization, driven by nesting on bipartite lattices, is associated to a potential energy gain) to a Heisenberg AF (whose stabilization, driven by the onset of coherence between preformed local moments, is associated to a kinetic energy gain). We notice that this evolution is not specific to the AF solution, but it is expected to be quite general. For instance, it is found also for the superconducting phase of the attractive Hubbard model, where it corresponds to the BCS-BEC crossover \cite{Toschi2005sc}.

Finally, from an algorithmic perspective, the information which could be gained from these investigations might provide crucial hints for the future development and the applicability of bold diagrammatic resummation schemes and/or of diagrammatic extensions \cite{rohringer2018} of DMFT or Two-particle Self-Consistent theory \cite{vilk1997} in the SU(2)-broken-symmetry phases \cite{delre2021su2,delre2024}.

For these reasons, in this paper, we extend the previous studies of the manifestations of the perturbation theory breakdown to the AF-ordered phase. To this aim, we have systematically performed two-particle DMFT calculations in the low-temperature AF-ordered phase of the half-filled Hubbard model. These calculations allow for a thorough investigation of possible divergences appearing in the irreducible vertex functions, as well as to study their relation with specific physical properties which characterize, respectively, the Slater and the Heisenberg AF regime. 
For a comparative analysis of the non-perturbative effects in the symmetry-broken phase w.r.t.~approximation schemes, which rely on the self-consistent perturbation expansion, we integrate our DMFT study with corresponding calculations in conventional static mean-field theory and the random phase approximation (MF/RPA).

The paper is structured as follows: In \cref{methods}, we define the specific Hubbard model, as well as the necessary one- and two-particle quantities for our DMFT and MF/RPA calculations. \cref{DMFT Results} shows our DMFT results for the divergences of the irreducible vertex in the AF phase and investigates their appearance in connection to the different behavior of the local physical charge response in the Slater and Heisenberg AF. \cref{AFbse} takes an eigenvalue perspective on the Bethe-Salpeter equation in the AF phase and discusses possible consequences for phase-separation instabilities. Conclusion and outlook of the paper are presented in Section \ref{Conclusion} and \ref{Outlook}, respectively.

\section{Formalism and Methods}
\label{methods}
\subsection{Bipartite Hubbard Model in Mean-Field and Dynamical Mean-Field Theory}

\begin{figure}[b!]
    \centering
    \begin{tikzpicture}[scale=0.8]
        \begin{feynhand}
            \draw [dashed,gray] (-1.9,-1.15) -- (-0.75,0);
            \draw [dashed,gray] (-1.9,1.15) -- (-0.75,0);
            \draw [dashed,gray] (3.4,-1.15) -- (3.2,-0.95);
            \draw [dashed,gray] (3.4,1.15) -- (3.2,0.95);
            \draw [dashed,gray] (2.8,-0.55) -- (2.25,0);
            \draw [dashed,gray] (2.8,0.55) -- (2.25,0);
            \draw [dashed,gray] (0.35,-1.9) -- (0.75,-1.5);
            \draw [dashed,gray] (1.15,-1.9) -- (0.75,-1.5);
            \draw [dashed,gray] (0.35,1.9) -- (0.75,1.5);
            \draw [dashed,gray] (1.15,1.9) -- (0.75,1.5);
            \draw [gray] (-0.75,0) -- (0.75,1.5);
            \draw [gray] (-0.75,0) -- (0.75,-1.5);
            \draw [gray] (0.75,1.5) -- (2.25,0);
            \draw [gray] (0.75,-1.5) -- (2.25,0);
            \draw[fill=colorA] (0,0) circle (0.2);
            \draw[fill=colorA] (-1.5,-1.5) circle (0.2);
            \draw[fill=colorA] (-1.5,1.5) circle (0.2);
            \draw[fill=colorA] (1.5,-1.5) circle (0.2);
            \draw[fill=colorA] (1.5,1.5) circle (0.2);
            \draw[fill=colorA] (3,0) circle (0.2);
            \draw[fill=colorB] (-1.5,0) circle (0.2);
            \draw[fill=colorB] (1.5,0) circle (0.2);
            \draw[fill=colorB] (0,-1.5) circle (0.2);
            \draw[fill=colorB] (0,1.5) circle (0.2);
            \draw[fill=colorB] (3,1.5) circle (0.2);
            \draw[fill=colorB] (3,-1.5) circle (0.2);
            \node at (0,-0.4) {$A$};
            \node at (1.5,-0.4) {$B$};
            \node at (2.5,2) {$t \, \e^{\ii(k_x + k_y)}$};
            \propag[plain, with reversed arrow=0] (0.2,0.2) to [quarter left, edge label =$t$] (1.3,0.2);
            \propag[plain, with arrow=0.99] (1.7,0.2) to [quarter left, edge label =$\quad \quad \quad t \, \e^{\ii 2 k_x }$] (2.8,0.2);
            \propag[plain, with reversed arrow=0] (1.7,1.3) to [quarter left] (1.7,0.2);
            \propag[plain, with arrow=0.99] (1.7,-0.2) to [quarter left, edge label = $t \, \e^{\ii (k_x - k_y)}$] (1.7,-1.3);
            \draw[->] (-1.9,-1.9) -- (-1.9,-0.7);
            \draw[->] (-1.9,-1.9) -- (-0.7,-1.9);
            \draw [lightgray] (0.75,-0.1) -- (0.75,0.1);
            \draw [lightgray] (0.65,0) -- (0.85,0);
            \node at (-2.2,-1.) {$y$};
            \node at (-1.,-2.2) {$x$};
            \draw [dashed,gray] (4.5,-1.9) -- (4.5,1.9);
            \draw [dashed,gray] (4.5,-1.9) -- (8.3,-1.9);
            \draw [dashed,gray] (4.5,1.9)  -- (8.3,1.9);
            \draw [dashed,gray] (8.3,-1.9) -- (8.3,1.9);
            \draw [gray] (4.5,0) -- (6.4,1.9);
            \draw [gray] (4.5,0) -- (6.4,-1.9);
            \draw [gray] (8.3,0) -- (6.4,1.9);
            \draw [gray] (8.3,0) -- (6.4,-1.9);
            \draw[->] (4.5,-1.9) -- (4.5,-0.7);
            \draw[->] (4.5,-1.9) -- (5.7,-1.9);
            \draw[->] (4.5,0) -- (5.3485,0.8485);
            \draw[->] (4.5,0) -- (5.3485,-0.8485);
            \draw [lightgray] (6.4,-0.1) -- (6.4,0.1);
            \draw [lightgray] (6.3,0) -- (6.5,0);
            \node at (4.2,-1.) {$k_y$};
            \node at (5.4,-2.2) {$k_x$};
            \node at (4.8,0.8) {$k'_y$};
            \node at (5.2,-0.3) {$k'_x$};
            \node at (4.1,-1.7) {$-\pi$};
            \node at (4.2,1.8) {$\pi$};
            \node at (4.5,-2.2) {$-\pi$};
            \node at (8.1,-2.2) {$\pi$};
        \end{feynhand}
    \end{tikzpicture}
    \caption{Left: Bipartite Hubbard model on the square lattice. In the Fourier transform the hopping parameters to the next neighbors acquire a phase factor depending on the distance of their unit cells (squares). Right: Reduced Brillouin zone (RBZ) of the bipartite model.}
    \label{fig:lattice}
\end{figure}
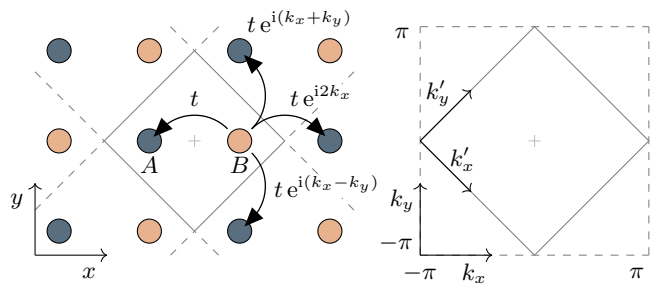

We study a bipartite square lattice Hubbard model with nearest neighbor hopping $t=1/4$, whose Hamiltonian reads
\begin{equation}
\begin{split}
   H =& \sum_{\vk\sigma} \left(\create{A\vk\sigma}, \,  \create{B\vk\sigma}\right) 
   \begin{pmatrix}
            -\mu & \varepsilon_{\vk}\\ 
            \varepsilon_{\vk}^* & - \mu
   \end{pmatrix}
   \begin{pmatrix}
    \annihil{A\vk\sigma} \\  \annihil{B\vk\sigma}
    \end{pmatrix}\\ 
    &+ U \sum_{i\alpha} \hat{n}_{\alpha i\up} \hat{n}_{\alpha i\down},
\end{split}
\end{equation}
where $\mu$ is the chemical potential, $U$ the Hubbard interaction, $\hat{n}_{\alpha i\sigma}=\create{\alpha i \sigma} \annihil{\alpha i \sigma}$ the occupation of spin $\sigma$ on site $\alpha=(A,B)$ in the $i$th unit cell, $\ancreate{\alpha \vk \sigma}$ the corresponding Fourier transform, and $\varepsilon_{\vk} = -2 t\, \e^{\ii k_x} (\cos{k_x}+\cos{k_y})$ the dispersion relation (see \cref{fig:lattice}), where $\vk$ runs over the reduced Brillouin zone (RBZ)\footnote{We choose the basis of the bipartite lattice such that the dispersion remains periodic within the reduced Brillouin zone~\cite{girvin2019}.}: $k_x = k'_x+k'_y$ and $k_y = k'_y-k'_x$, with $k'_x,k'_y \in [(-\pi/2,-\pi/2), (\pi/2,\pi/2))$.

At particle-hole (\ph) symmetry $\mu\!=\!U/2$, which corresponds to half-filling $n\!=\!n_{\alpha\up}+n_{\alpha\down}\!=\langle \hat{n}_{\alpha\up}+\hat{n}_{\alpha\down}\rangle\!=\!1$, and zero temperature, the model is expected to order antiferromagnetically with a non zero staggered magnetization $m = m_A=n_{A\up}-n_{A\down}=-m_B$ on the two lattice sites. Without loss of generality, such magnetization (if present) can be oriented along the $z$-axis, and the corresponding Dyson equation then reads
\begin{align}
    &G_{k\sigma} =
    \begin{pmatrix}
            G^{AA}_{k\sigma} &  G^{AB}_{k\sigma}\\
             G^{BA}_{k\sigma}&  G^{BB}_{k\sigma}
   \end{pmatrix}\\ \nonumber
    &= \frac{1}{\zeta_{A k\sigma} \zeta_{B k\sigma} - \abs{\varepsilon_\vk+\Sigma_{ABk\sigma}}^2}
    \begin{pmatrix}
           \zeta_{B k\sigma} & \varepsilon_{\vk} +\Sigma_{ABk\sigma}\\ 
            \varepsilon_{\vk}^*+\Sigma^*_{ABk\sigma} & \zeta_{A k\sigma}
   \end{pmatrix},
\end{align}
where $\zeta_{\alpha k \sigma} = \ii\nu + \mu -\Sigma_{\alpha k\sigma}$, the fermionic Matsubara frequencies are defined $\nu=(2\mathrm{n}+1) \pi T$, with $\mathrm{n}\in \mathbb{Z}$, and $\Sigma_{\alpha k\sigma}$ is the electronic self-energy on site $\alpha$ and $\Sigma_{ABk\sigma}=\Sigma^*_{BAk\sigma}$ between site $A$ and $B$, with $k=(\nu,\vk)$.

From the symmetry $(A,\sigma) \leftrightarrow (B,-\sigma)$ in the antiferromagnetic (AF) phase, it follows that $\zeta_{B k \sigma} \!=\! \zeta_{A k -\sigma}$. In the paramagnetic (PM) phase, when $\zeta_{\alpha k \sigma} \!=\! \zeta_{\alpha k -\sigma}$, diagonalizing the matrix $G_{ k\sigma}$ returns two Green's functions 
\begin{equation}
    \mathcal{G}_{k\sigma \pm} = \frac{1}{\zeta_{k\sigma} \pm \abs{\varepsilon_\vk +\Sigma_{ABk\sigma}}}
\end{equation}
associated to the two ``bands'' in the RBZ, corresponding to $\vk$ and $\vk+(\pi,\pi)$ of the full Brillouin zone.

In the course of the paper, we study the half-filled Hubbard model within two self-consistent approximate methods, namely (i) the perturbative  conventional static mean-field theory and random phase approximation 
 (MF/RPA) approach and (ii) the non-perturbative dynamical mean-field theory (DMFT). By means of the latter, we directly extend the study of non-perturbative effects reported in recent literature \cite{reitner2020,chalupa2021,reitner2024,adler2024,kowalski2024} to the symmetry-broken phase. 
It is important to recall that, while being conserving approximations \cite{bickers2004}, both RPA and DMFT do not fulfill the Mermin-Wagner theorem, yielding an antiferromagnetic ordered phase at finite temperatures in two dimensions. 
Our two-dimensional (2D) results are expected to be qualitatively similar to those of a three-dimensional (3D) Hubbard model, where an actual AF long-range order is expected also at finite temperature. Such an agreement could even turn quantitative if we scale the hopping in order to have the same variance of the non-interacting density of states in two and three dimensions\footnote{For instance, this can be realized by taking $t=1/4$ for a 2D square lattice  (as done here) and  $t = 1/(2\sqrt{6})$ for a 3D cubic lattice~\cite{georges1996}.}.

Due to such a fairly good equivalence, at the DMFT
level,  we decided to perform our calculations in two dimensions. This allows us to better match the recent literature developments~\cite{bonetti2022,goremykin2024}, which restore the Mermin-Wagner theorem starting from a symmetry-broken (dynamical) mean-field solution, as well as to provide a reference for future non-perturbative studies of generalized two-particle susceptibilities and vertex functions of 2D systems beyond the local DMFT realm. We will elaborate on the plausible consequences of our results for the Hubbard model in two dimensions in \cref{Outlook}.

In both methods, MF/RPA and DMFT, the self-energy  only contributes on-site $\Sigma_{ABk\sigma}=0$ and $\Sigma_{\alpha k\sigma}$ becomes independent of $\vk$: (i) in MF/RPA  $\Sigma_{\alpha k\sigma}\approx (U/2) (n- m_\alpha)$, and (ii) in DMFT where $\Sigma_{\alpha k\sigma}\! \approx\! \Sigma_{\alpha\nu\sigma}$ \cite{georges1996}. For the latter $\Sigma_{\alpha\nu\sigma}$ corresponds to the on-site self-energy calculated from an auxiliary Anderson impurity model (AIM), which in our case is solved using the continuous-time quantum Monte Carlo \cite{gull2011}  {\sl{w2dynamics}}~\cite{wallenberger2019} software package.

In the MF approximation (i), the self-consistency is achieved from iterated calculations of the static staggered magnetization~\cite{Chubukov1992}
\begin{equation}
    m_\alpha = \frac{1}{V\beta}\sum_k \left( G^{\alpha\alpha}_{k \up} - G^{\alpha\alpha}_{k \down} \right)\e^{\ii \nu 0^+},
\end{equation}
where $V=\mathfrak{V}/2$ is half of the volume $\mathfrak{V}$ and $\beta=1/T$ the inverse temperature.
While in DMFT (ii), the condition of self-consistency is attained from the calculation of the (frequency-dependent) local Green's function~\cite{georges1996}
\begin{equation}
    G^{\alpha\alpha}_{\nu\sigma} = \frac{1}{V}\sum_\vk G^{\alpha\alpha}_{k \sigma} = \frac{1}{\zeta_{\alpha\nu\sigma}-\Delta_{\alpha\nu\sigma}},
\end{equation}
where $\Delta_{\alpha\nu\sigma}$ is the hybridization function of the AIM.

\subsection{Generalized Susceptibilities}
\label{sus}
Information about the physical response and fluctuations in charge and spin degrees of freedom of the model is encoded at the two-particle level in the generalized susceptibilities~\cite{rohringer2012,rohringer2018,kugler2021multipoint} defined as:
\begin{gather}
\label{eq:chi}
\chi^{\alpha\alpha'kk'}_{q\sigma\sigma'} = \int d\tau_1 d\tau_2 d\tau_3 \,
e^{-i\nu\tau_1} e^{i(\nu+\omega)\tau_2} e^{- i(\nu'+\omega)\tau_3}  \\ \nonumber
     \times  \left[ \langle \mathcal{T} \create{\alpha\vk\sigma}(\tau_1) \annihil{ \alpha\vk+\vq\sigma }(\tau_2)
    \create{\alpha'\vk'+\vq\sigma' }(\tau_3) \annihil{\alpha'\vk'\sigma'}(0)\rangle  \right. \\ \nonumber
         - \langle \mathcal{T} \left.\create{\alpha\vk\sigma}(\tau_1)
   \annihil{ \alpha\vk+\vq\sigma }(\tau_2) \rangle
     \langle \mathcal{T}   \create{\alpha'\vk'+\vq\sigma' }(\tau_3) \annihil{\alpha'\vk'\sigma'}(0)\rangle \right], 
\end{gather}
where $\sigma, \sigma'$ denote the spins on sites $\alpha, \alpha'$ with fermionic $k\!=\!(\nu,\mathbf{k}), k'\!=\!(\nu',\mathbf{k'})$  and bosonic $q=(\omega,\mathbf{q})$ ($ \omega\!=\! 2\mathrm{n} \pi T $, $\mathrm{n}\!\in\!\mathbb{Z}$) four-momenta, $\mathcal{T}$ denotes the imaginary time ordering operator, and $\big< \dots \big> = (1/Z)\Tr(\e^{-\beta H} \dots)$ the thermal expectation value.  To shorten the notation, we define the generalized susceptibilities corresponding to the momentum-dependent response of our square lattice system as
\begin{align}  \bchi^{\alpha\alpha'}_{\vq\sigma\sigma'}(\omega) &\coloneqq \frac{1}{V^2}\sum_{\vk\vk'}\chi^{\alpha\alpha'kk'}_{q\sigma\sigma'},
\end{align}
i.e.~as matrices in the fermionic frequency space $\nu, \nu'$.
Further, for quantities corresponding to the static/isothermal $\omega\!=\!0$ response \cite{watzenboeck2022}, we will omit the corresponding zero transfer frequency argument $\bchi^{\alpha\alpha'}_{\vq\sigma\sigma'}\coloneqq\bchi^{\alpha\alpha'}_{\vq\sigma\sigma'}(0)$.
The local generalized susceptibilities $\bchi^{\alpha\alpha'}_{\sigma\sigma'}(\omega)$ (in MF/RPA) can be then obtained by summing over all momenta $\vq$ 
\begin{equation}    \bchi^{\alpha\alpha'}_{\sigma\sigma'}(\omega) = \frac{1}{V}\sum_{\vq} \bchi^{\alpha\alpha'}_{\vq\sigma\sigma'}(\omega),
\end{equation}
which in the limit of infinite lattice connectivity, where DMFT  becomes rigorously exact \cite{metzner1989,georges1996}, would also correspond to the on-site generalized susceptibilities $\bchi_{\sigma\sigma'} = \bchi^{AA}_{\sigma\sigma'}=\bchi^{BB}_{-\sigma-\sigma'}$, and $\bchi^{\alpha\alpha'}_{\sigma\sigma'}=\delta_{\alpha\alpha'}\bchi^{\alpha\alpha}_{\sigma\sigma'}$ of the auxiliary AIM of the self-consistent DMFT solution. 
In the two-dimensional case considered here, where DMFT is an approximation, the generalized susceptibilities of the auxiliary AIM will be used for the calculations of the corresponding local quantities. 

\begin{figure}[t]
    \centering
    \includegraphics[width=\textwidth]{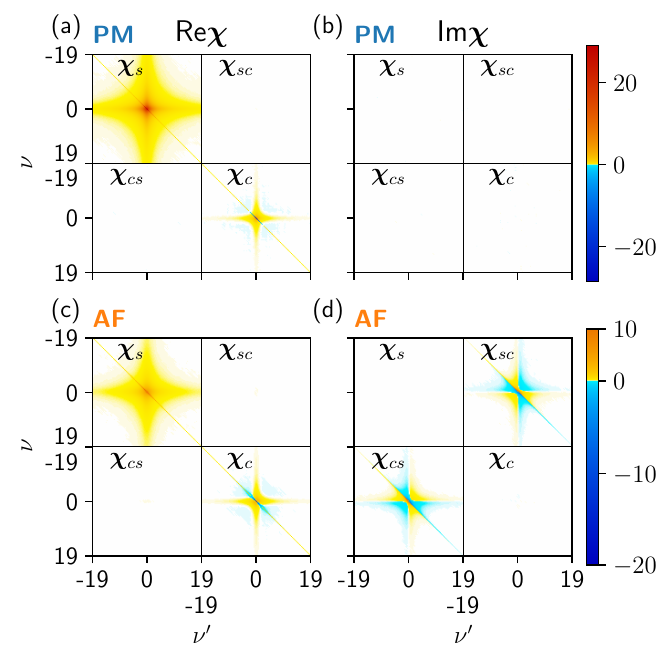}
    \caption{Real (left column) and imaginary part (right column) of the local generalized susceptibility $\bchi$ calculated in DMFT for a Hubbard interaction value of $U=3$, and temperature $T=1/20$ at particle-hole symmetry ($\mu=U/2$). The top row shows the paramagnetic (PM), and the bottom row the corresponding antiferromagnetic (AF) solution.}
    \label{fig:matrix}
\end{figure}

We also recall that, for non-interacting systems as well as when all two-particle scattering processes (i.e., the so-called vertex corrections) are neglected, the susceptibility expressions above reduce to the corresponding ``bubble" terms,  defined as
\begin{equation}  \bchi^{\alpha\alpha'}_{0\vq\sigma\sigma'}(\omega) = -\frac{\beta}{V}\sum_{\vk} \delta_{\nu\nu'} \delta_{\sigma\sigma'}G^{\alpha\alpha'}_{k\sigma}G^{\alpha'\alpha}_{k+q\sigma},
\end{equation}
and, for the on-site counterpart, as
\begin{equation}
\begin{split}
    \bchi_{0\sigma\sigma'}(\omega) &= \frac{1}{V} \sum_\vq \bchi^{AA}_{0\vq\sigma\sigma'}(\omega)\\
    &= \frac{1}{V} \sum_\vq \bchi^{BB}_{0\vq-\sigma-\sigma'}(\omega).
\end{split}
\end{equation}
To obtain the local generalized susceptibilities in the usual charge ($c$) and spin ($s$) notation 
\begin{equation}
\label{eq:loc_chi}
    \bchi = 
    \begin{pmatrix}
        \bchi_s & \bchi_{sc}\\
        \bchi_{cs} & \bchi_c,
    \end{pmatrix}
\end{equation}
which is directly associated to the local density and magnetic linear response of the system,
we use the following definitions (cf. also \cite{delre2021su2})
\begin{align}
\label{eq:loc_chi_s}
     \bchi_{ s} &= \frac{1}{2} (\bchi_{\up\up}+\bchi_{\down\down}-\bchi_{\up\down}-\bchi_{\down\up}),\\
     \bchi_{ sc} &= \frac{1}{2} (\bchi_{ \up\up}-\bchi_{ \down\down}+\bchi_{ \up\down}-\bchi_{ \down\up}),\\
     \bchi_{ c} &= \frac{1}{2} (\bchi_{ \up\up}+\bchi_{ \down\down}+\bchi_{ \up\down}+\bchi_{ \down\up}),\\
     \bchi_{ cs} &= \frac{1}{2} (\bchi_{ \up\up}-\bchi_{ \down\down}-\bchi_{ \up\down}+\bchi_{ \down\up}).
\end{align}
For the corresponding bubble terms we define
\begin{align}
    \bchi_{0} &= \frac{1}{2} (\bchi_{0\up\up}+\bchi_{0\down\down}),\\
    \overline{\bchi}_{0} &= \frac{1}{2} (\bchi_{0\up\up}-\bchi_{0\down\down}).
\end{align}
Before we turn to illustrate the formalism of the Bethe-Salpeter equation, it is useful to shortly highlight some general properties of $\bchi$ as a matrix in fermionic Matsubara frequency and charge-spin space.
First of all, it is evident that in the SU(2)-symmetric PM phase, where $\bchi_{\sigma\sigma'}=\bchi_{-\sigma-\sigma'}$, the mixed charge and spin components vanish $\bchi_{ sc}=\bchi_{ cs} = 0$. In general, at \ph-symmetry (half-filling) $\bchi_{s}$, $\bchi_{c}$ become real and bisymmetric matrices~\cite{rohringer2012,springer2020,essl2024}, while $\bchi_{sc}$, $\bchi_{cs}$ become purely imaginary centro-Hermitian matrices~\cite{lee1980,reitner2024} (cf. \cref{fig:matrix}). Further, by exchanging $\nu \leftrightarrow \nu'$ (for $\omega=0$), corresponding to a matrix transpose, we see that  $\bchi$ is  a symmetric matrix $\bchi^T=\bchi$ (for details see \cref{app:chi_symm}).

From these properties, it follows that while in the PM phase the eigenvalues $\lambda_i$ of $\bchi$ are purely real, in the AF phase they can be either real or appear in complex-conjugate pairs. Further, if $\bchi$ is diagonalizable\footnote{In principle,  $\bchi$ can become defective, and hence, not diagonalizable, at specific ``exceptional points'' in the phase-diagram. For a detailed discussion we refer the reader to Ref.~\cite{reitner2024,essl2024}.}, we find an orthonormal eigenbasis $\mathbf{v}_i$, where $\bchi \mathbf{v}_i = \lambda_i\mathbf{v}_i$, with $\mathbf{v}^T_i \mathbf{v}_j=\delta_{ij}$. The corresponding eigenvectors $\mathbf{v}_i=(\bm{v}^s_i,\bm{v}^c_i)$ are either symmetric in Matsubara frequencies in the spin component $\bm{v}^s_i$ and anti-symmetric in the charge component $\bm{v}^c_i$ or vice versa (this holds also in the AF phase, for a proof refer to \cref{app:v_symm}). Since for the calculation of the physical response, we sum over the Matsubara frequencies $(1/\beta)\sum_\nu \mathbf{v}_i$ (see \cref{sec:bse}), eigenvalues contribute solely to the physical charge or spin response, as the anti-symmetric components vanish. 

\subsection{Static Response and Bethe-Salpeter Equation}
\label{sec:bse}
\noindent

When calculating the static physical responses of our system,  we must recall that in a bipartite lattice, the $\vq$-dependence of the density and magnetic operators 
\begin{align}
    \hat{n}_\alpha(\vq) &= \sum_{\sigma} \hat{n}_{\alpha\sigma}(\vq) = \frac{1}{V} \sum_{\alpha\vk\sigma} \create{\alpha\vk\sigma}\annihil{\alpha\vk+\vq\sigma},\\
\hat{m}_\alpha(\vq) &= \hat{n}_{\alpha\up}(\vq) - \hat{n}_{\alpha\down}(\vq)
\end{align}
becomes periodic in the RBZ $[(-\pi/2,-\pi/2), (\pi/2,\pi/2))$. The full $\fq$-dependence in the Brillouin zone (BZ) $[(-\pi,-\pi), (\pi,\pi))$ of the operators in the single atomic basis 
\begin{align}
    \hat{n}(\fq) &= \sum_{\sigma} \hat{n}_\sigma(\fq)
    = \frac{1}{\mathfrak{V}} \sum_{\vk'\sigma} \create{\vk'\sigma}\annihil{\vk'+\fq\sigma} ,\\
    \hat{m}(\fq) &= \hat{n}_\up(\fq) - \hat{n}_\down(\fq),
\end{align}
where $\mathfrak{V}=2V$ and $\vk'$ runs here over the full BZ, can be restored by the two ``band'' expressions in the bipartite basis, corresponding to $\fq=\vq$ and $\fq=\vq+(\pi,\pi)$  of the single atomic basis~\cite{delre2021su2}:
\begin{align}
    \hat{n}(\vq) &= \frac{1}{2}(\hat{n}_A(\vq) + \hat{n}_B(\vq)),\\
    \hat{n}(\vq+(\pi,\pi)) &= \frac{1}{2}(\hat{n}_A(\vq) - \hat{n}_B(\vq)),\\
    \hat{m}(\vq) &= \frac{1}{2}(\hat{m}_A(\vq) + \hat{m}_B(\vq)),\\
    \hat{m}(\vq+(\pi,\pi)) &= \frac{1}{2}(\hat{m}_A(\vq) - \hat{m}_B(\vq)).
\end{align}
The static physical responses $\partial n(\fq)/\partial \tilde{\mu}$, $\partial n(\fq)/\partial h$,  $\partial m(\fq+(\pi,\pi))/\partial \tilde{\mu}$, and $\partial m(\fq+(\pi,\pi))/\partial h$ to the external perturbations  $\tilde{\mu}$ and $h$ (at $\tilde{\mu}=0$ and $h=0$), which couple to $ n(-\fq)$ and $m(-\fq- (\pi,\pi))$, respectively, can then be calculated from the generalized susceptibilities~\cite{delre2021su2}\footnote{Note that the responses $\partial n(\fq\! +\!(\pi,\pi))/\partial h= \partial[n_A(\vq) \mp n_B(\vq)]/\partial h$ and $\partial m(\fq )/\partial \tilde{\mu}= \partial [m_A(\vq) \pm m_B(\vq)]/\partial \tilde{\mu}$ vanish due to the AF symmetry $(A,\sigma) \leftrightarrow (B,-\sigma)$.}$^{,}$\footnote{We note, in passing, that through the Ward-Identities~\cite{ward1950, krien2017} the eigenvectors of the generalized susceptibility  $\bchi_{\vq=0\pm}$ in \cref{eq:BSE} can be also regarded as spectral decomposition of the changes in the Green's function, e.g.~$ (1/2)\partial[\sum_\sigma (G^{AA}_{\nu\sigma}\pm G^{BB}_{\nu\sigma})]/\partial\tilde{\mu}$, $(1/2)\partial[\sum_\sigma (G^{AA}_{\nu\sigma}\pm G^{BB}_{\nu\sigma})]/\partial h$~\cite{kowalski2024}.}:
\begin{align}
\label{eq:dndmu}
    \frac{1}{2}\pdv{\left(n_A(\vq) \pm n_B(\vq)\right)}{\tilde{\mu}} &= \frac{1}{\beta^2} \sum_{\nu\nu'} \left(\bchi_{\vq c}^{AA} \pm \bchi_{\vq c}^{AB} \right),\\
\label{eq:dndh}
    \frac{1}{2}\pdv{\left(n_A(\vq) \pm n_B(\vq)\right)}{h} &= \frac{1}{\beta^2} \sum_{\nu\nu'} \left(\bchi_{\vq cs}^{AA} \mp \bchi_{\vq cs}^{AB} \right),\\
\label{eq:dmdmu}
   \frac{1}{2}\pdv{\left(m_A(\vq) \mp m_B(\vq)\right)}{\tilde{\mu}} &= \frac{1}{\beta^2} \sum_{\nu\nu'} \left(\bchi_{\vq sc}^{AA} \pm \bchi_{\vq sc}^{AB} \right),\\
\label{eq:dmdh}
    \frac{1}{2}\pdv{\left(m_A(\vq) \mp m_B(\vq)\right)}{h} &= \frac{1}{\beta^2} \sum_{\nu\nu'} \left(\bchi_{\vq s}^{AA} \mp \bchi_{\vq s}^{AB} \right).
\end{align}
These are obtained from the following Bethe-Salpeter equation (BSE)~\cite{delre2021su2} (see also \cref{app:bse}):
\begin{equation}
    \label{eq:BSE}
    \begin{split}
    \bchi_{\vq \pm} =& \begin{pmatrix}
        \bchi_{\vq s}^{AA} \pm \bchi_{\vq s}^{AB} & \bchi_{\vq sc}^{AA} \mp \bchi_{\vq sc}^{AB} \\
         \bchi_{\vq cs}^{AA} \pm \bchi_{\vq cs}^{AB} & \bchi_{\vq c}^{AA} \mp \bchi_{\vq c}^{AB} 
    \end{pmatrix} \\
    =& \left[\Gamma + \bchi^{-1}_{0\vq \pm}\right]^{-1},
    \end{split}
\end{equation}
where $\bchi^{-1}_{0\vq \pm}$ is the corresponding ``bubble'' susceptibility and $\Gamma$ is the irreducible vertex. 

The former term is explicitly defined as:
\begin{equation}
     \bchi_{0\vq \pm} = 
     \begin{pmatrix}
       \bchi_{0\vq}^{AA} \pm \bchi_{0\vq}^{AB} & \overline{\bchi}_{0 \vq}^{AA} \mp \overline{\bchi}_{0\vq}^{AB} \\
        \overline{\bchi}_{0\vq}^{AA} \pm \overline{\bchi}_{0\vq}^{AB} & \bchi_{0\vq}^{AA} \mp \bchi_{0\vq}^{AB} 
     \end{pmatrix}.
\end{equation}
Further, we used the following  notations:
\begin{align}
\label{eq:sus_s}
     \bchi^{\alpha\alpha'}_{\vq s} &= \frac{1}{2} (\bchi^{\alpha\alpha'}_{\vq\up\up}+\bchi^{\alpha\alpha'}_{\vq\down\down}-\bchi^{\alpha\alpha'}_{\vq\up\down}-\bchi^{\alpha\alpha'}_{\vq\down\up}),\\
\label{eq:sus_sc}
     \bchi^{\alpha\alpha'}_{\vq sc} &= \frac{1}{2} (\bchi^{\alpha\alpha'}_{\vq \up\up}-\bchi^{\alpha\alpha'}_{\vq \down\down}+\bchi^{\alpha\alpha'}_{\vq \up\down}-\bchi^{\alpha\alpha'}_{\vq \down\up}),\\
\label{eq:sus_c}
     \bchi^{\alpha\alpha'}_{\vq c} &= \frac{1}{2} (\bchi^{\alpha\alpha'}_{\vq \up\up}+\bchi^{\alpha\alpha'}_{\vq \down\down}+\bchi^{\alpha\alpha'}_{\vq \up\down}+\bchi^{\alpha\alpha'}_{\vq \down\up}),\\
\label{eq:sus_cs}
     \bchi^{\alpha\alpha'}_{\vq cs} &= \frac{1}{2} (\bchi^{\alpha\alpha'}_{\vq \up\up}-\bchi^{\alpha\alpha'}_{\vq \down\down}-\bchi^{\alpha\alpha'}_{\vq \up\down}+\bchi^{\alpha\alpha'}_{\vq \down\up}),\\
      \bchi^{\alpha\alpha'}_{0\vq } &= \frac{1}{2} (\bchi^{\alpha\alpha'}_{0\vq \up\up}+\bchi^{\alpha\alpha'}_{0\down\down}),\\
       \overline{\bchi}^{\alpha\alpha'}_{0\vq } &= \frac{1}{2} (\bchi^{\alpha\alpha'}_{0\vq \up\up}-\bchi^{\alpha\alpha'}_{0\vq \down\down}),
\end{align}
where the local generalized susceptibilities are denoted by omitting the $\vq$ index.
We note here that, due to the spontaneous breaking of the SU(2)-symmetry associated to the AF-ordering in the BSE of \cref{eq:BSE},  the charge and the spin response of the two different ``bands'' (corresponding to $\vq$ and $\vq+(\pi,\pi)$) always couple to each other.

As for the irreducible vertex $\Gamma$, in the random phase approximation (MF/RPA), it simply reads
\begin{equation}
\label{eq:gamma_rpa}
    \Gamma = \frac{1}{\beta^2}
    \begin{pmatrix}
        -U & 0  \\
         0 & U
    \end{pmatrix},
\end{equation}
while, in DMFT, it is formally defined as
\begin{equation}
\label{eq:gamma_dmft}
    \Gamma = 
    \begin{pmatrix}
        \bchi_{s} & \bchi_{sc}  \\
         \bchi_{cs} & \bchi_{c}
    \end{pmatrix}^{-1}
    - 
     \begin{pmatrix}
        \bchi_{0} & \overline{\bchi}_{0}  \\
        \overline{\bchi}_{0} & \bchi_{0}
    \end{pmatrix}^{-1}.
\end{equation}
By taking a look at \cref{eq:gamma_dmft}, we can readily note that the irreducible vertex  $\Gamma$ of DMFT may actually diverge, with the corresponding BSE becoming non-invertible if $\bchi$ has a zero eigenvalue $\lambda_i=0$ \cite{schafer2013,schafer2016}. In this case, as it is clearly discussed in the literature \cite{kozik2015,gunnarsson2017,vucicevic2018}, techniques relying on the self-consistent resummation of a perturbative series of bold diagrams for $\Sigma[G]$ start to fail.
In particular, the breakdown of self-consistent perturbation theory can be directly ascribed to the occurrence of the crossing between a physical and an unphysical solution~\cite{kozik2015,stan2015,tarantino2018,vucicevic2018,kim2020} of the Luttinger Ward functional, $\Phi[G]$, which directly implies~\cite{gunnarsson2017}, the divergence of the irreducible vertex functions~\cite{schafer2013,schafer2016,springer2020}. 
 In fact, such a relation can also be understood, considering that the irreducible vertex is formally defined through the respective functional derivative of the self-energy functional $\Gamma = \beta \delta\Sigma[G]/\delta G$~\cite{kozik2015,gunnarsson2017}, i.e.~the second functional derivative of $\Phi[G]$. The corresponding definition of $\Gamma$ in the bipartite basis explicitly reads:
\begin{equation}
\begin{split}
\Gamma = \frac{\beta}{2}
     &  \left(\begin{matrix}
        \left(
        \frac{\delta\Sigma^{AA}_{\nu\up}}{\delta G^{AA}_{\nu'\up}} 
        +\frac{\delta\Sigma^{AA}_{\nu\down}}{\delta G^{AA}_{\nu'\down}}
        -\frac{\delta\Sigma^{AA}_{\nu\up}}{\delta G^{AA}_{\nu'\down}}
        -\frac{\delta\Sigma^{AA}_{\nu\down}}{\delta G^{AA}_{\nu'\up}}
        \right)\\
        \left(
         \frac{\delta\Sigma^{AA}_{\nu\up}}{\delta G^{AA}_{\nu'\up}} 
        -\frac{\delta\Sigma^{AA}_{\nu\down}}{\delta G^{AA}_{\nu'\down}}
        -\frac{\delta\Sigma^{AA}_{\nu\up}}{\delta G^{AA}_{\nu'\down}}
        +\frac{\delta\Sigma^{AA}_{\nu\down}}{\delta G^{AA}_{\nu'\up}}
        \right)
        \end{matrix}\right.\\ 
    &   \phantom{\Biggl(}\left.\begin{matrix}
        \left(
         \frac{\delta\Sigma^{AA}_{\nu\up}}{\delta G^{AA}_{\nu'\up}} 
        -\frac{\delta\Sigma^{AA}_{\nu\down}}{\delta G^{AA}_{\nu'\down}}
        +\frac{\delta\Sigma^{AA}_{\nu\up}}{\delta G^{AA}_{\nu'\down}}
        -\frac{\delta\Sigma^{AA}_{\nu\down}}{\delta G^{AA}_{\nu'\up}}
        \right)\\
        \left(
         \frac{\delta\Sigma^{AA}_{\nu\up}}{\delta G^{AA}_{\nu'\up}} 
        +\frac{\delta\Sigma^{AA}_{\nu\down}}{\delta G^{AA}_{\nu'\down}}
        +\frac{\delta\Sigma^{AA}_{\nu\up}}{\delta G^{AA}_{\nu'\down}}
        +\frac{\delta\Sigma^{AA}_{\nu\down}}{\delta G^{AA}_{\nu'\up}}
        \right)
    \end{matrix}\right).
\end{split}
\end{equation}
Notably, a breakdown of the self-consistent perturbation theory has also been found in regimes where $\bchi$ shows only complex eigenvalues and $\Gamma$ is formally not divergent~\cite{vucicevic2018,gunnarsson2016,reitner2020,essl2024}. More recently, in such cases one precise condition\footnote{In particular, it has been demonstrated, that this condition controls the stability of the physical fixed-point of the iterative self-consistent perturbation theory, independently of the problem of the crossing of different branches of the Luttinger Ward functional.} limiting the applicability of self-consistent/iterative methods has been identified~\cite{essl2025} as a corresponding sign switch of the real part of an eigenvalue of
\begin{equation}
\label{eq:breakdown_cond}
\mathfrak{X}=
\begin{pmatrix}
        \bchi_{0} & \overline{\bchi}_{0}  \\
        \overline{\bchi}_{0} & \bchi_{0}
    \end{pmatrix}^{-1}
    \cdot
     \begin{pmatrix}
        \bchi_{s} & \bchi_{sc}  \\
         \bchi_{cs} & \bchi_{c}
    \end{pmatrix}_{.}
\end{equation}

At this point, it becomes convenient to also introduce the difference of the inverses of the two bubbles
\begin{align}
    &\Delta\bchi^{-1}_{0\vq} = \begin{pmatrix}\Delta\bchi^{-1}_{0\vq+} & \Delta\overline{\bchi}^{-1}_{0\vq-}\\
            \Delta\overline{\bchi}^{-1}_{0\vq+}& \Delta\bchi^{-1}_{0\vq-}
    \end{pmatrix} \\\nonumber
    &=\begin{pmatrix}
        \bchi^{AA}_{0\vq}+ \bchi^{AB}_{0\vq}& \overline{\bchi}^{AA}_{0\vq}- \overline{\bchi}^{AB}_{0\vq}\\\overline{\bchi}^{AA}_{0\vq}
        + \overline{\bchi}^{AB}_{0\vq} & \bchi^{AA}_{0\vq}- \bchi^{AB}_{0\vq}
    \end{pmatrix}^{-1}
    -
    \begin{pmatrix}
        \bchi^{AA}_0 & \overline{\bchi}^{AA}_0\\
        \overline{\bchi}^{AA}_0 & \bchi^{AA}_0
    \end{pmatrix}^{-1}.
\end{align}
This allows to  rewrite \cref{eq:BSE} with the help of \cref{eq:gamma_dmft} in a particularly insightful form for several DMFT applications:
\begin{equation}
    \label{eq:dmft_bse}
    \bchi_{\vq\pm} = \left[\begin{pmatrix}
        \bchi_{s} & \bchi_{sc}  \\
         \bchi_{cs} & \bchi_{c}
    \end{pmatrix}^{-1}
    +
    \begin{pmatrix}\Delta\bchi^{-1}_{0\vq\pm} & \Delta\overline{\bchi}^{-1}_{0\vq\mp}\\
            \Delta\overline{\bchi}^{-1}_{0\vq\pm}& \Delta\bchi^{-1}_{0\vq\mp}
    \end{pmatrix}
    \right]^{-1}.
\end{equation}
In fact, within DMFT, $\Delta\bchi^{-1}_{0\vq}$ can be regarded as the correction to the local generalized susceptibility $\bchi$ that is needed to obtain the lattice quantity $\bchi_{\vq\pm}$. The convenience of such rewriting becomes clear, if one observes that the $\vq$-dependence enters here only in $\Delta\bchi^{-1}_{0\vq}$.

\section{\label{DMFT Results}DMFT Results}
\subsection{Divergences of the Irreducible Vertex}
\begin{figure*}[ht]
    \centering
\includegraphics[width=\textwidth]{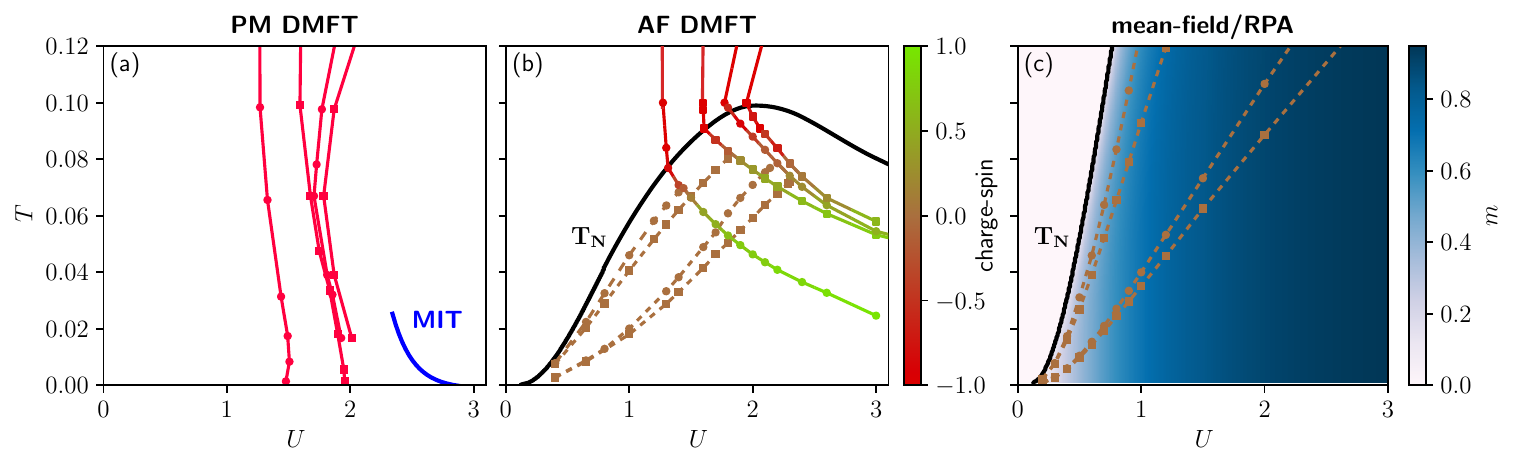}
     \caption{Location of the first four divergences of the two-particle irreducible vertex $\Gamma$ in the $T\! -\! U$ phase diagram for the paramagnetic (PM) DMFT solution in panel (a), the anti-ferromagnetic (AF) DMFT solution in panel (b), and the mean-field solution in panel (c) of the Hubbard model on a square lattice at half-filling. The divergences correspond to a zero-crossing of an eigenvalue $\lambda_i=0$ of the local generalized susceptibility $\bchi$ denoted as lines in the $T\! -\! U$ phase diagram. The color of the lines in panels (a) and (b) indicate the dominant contribution of the eigenvector to the charge (red) or spin (green) component of $\bchi$, while parameter sets where only $\Re\lambda=0$,   but $\Im\lambda \neq 0$ (i.e., no divergence of the vertex), are displayed as brown dashed lines.  Panel (a): The divergence lines in the PM phase have been reproduced from Ref.~\cite{schafer2016}. The solid blue line marks $U_{\text{crit}}$ of the para-magnetic Mott metal-to-insulator (MIT) transition, taken from Ref.~\cite{blumer2002,pelz2023}. Panel (b): The black solid line indicates the Néel temperature $T_N$ taken from Ref.~\cite{kunes2011}, where the low $U$ values have been estimated by the effective re-scaled $T_N$ of the mean-field solution based on Ref.~\cite{vandongen1994}, albeit with a slightly higher factor. Panel (c): The background color scale shows the magnetization.}
    \label{fig:div_lines}
\end{figure*}
Let us start by recalling the results of Refs.~\cite{schafer2013,schafer2016,pelz2023} which have investigated the divergences of the irreducible vertex $\Gamma$ of the DMFT solution for the PM phase of the square lattice Hubbard model at \ph-symmetry. The location of the first four reported divergences is displayed by the red lines in the $T\!-\!U$ phase diagram in panel (a) of \cref{fig:div_lines}. 

It should be underlined here that in the PM phase, as a result of the underlying SU(2)-symmetry, $\bchi$ becomes block-diagonal \cite{bickers2004}, resulting in fully separated charge and spin components $\bchi_s$ and $\bchi_c$. In this context \cite{schafer2013,schafer2016,pelz2023}, for positive $U$ values (i.e., the case of a repulsive Hubbard model, considered here),\footnote{The situation for $U<0$ at \ph-symmetry can be analytically mapped from $U>0$ by a Shiba transformation. For details we refer to Refs.~\cite{springer2020,essl2024}.} only $\bchi_c$\footnote{In fact, in the PM phase vertex divergences at \ph-symmetry and $U>0$ appear in the charge and in the particle-particle channel, but {\sl not} in the spin channel \cite{schafer2013,gunnarsson2016,schafer2016}.} shows eigenvalues that cross zero ($\lambda_i=0$) as a function of $U$, hence, only the charge component of $\Gamma$ diverges. In particular, with increasing interaction values $U$, first an eigenvalue $\lambda_i$ of $\bchi_c$ corresponding to an eigenvector $\mathbf{v}_i$ that is anti-symmetric (in $\bm{v}^c_i$, with $\bm{v}^s_i=0$) vanishes, while $\Gamma$ diverges (illustrated as red solid line with circles), and becomes negative for any higher $U$ value. By further increasing $U$ a second eigenvalue of $\bchi_c$, corresponding to a symmetric eigenvector, crosses zero (red solid line with squares). Eventually, for higher and higher values of $U$, more eigenvalues of $\bchi$ cross the zero line, with alternating  symmetry of the eigenvector (red solid lines with circles for anti-symmetric, squares for symmetric eigenvectors in $\bm{v}^c_i$). The resulting divergence lines in the $T\!-\!U$ phase diagram become denser with increasing $U$ towards the Mott metal-to-insulator transition (MIT, shown as blue line), which indeed represents an accumulation point of vertex divergences in the DMFT phase diagram \cite{pelz2023}.

Panel (b) of \cref{fig:div_lines} illustrates, instead, how the situation changes when AF ordering is allowed in our DMFT calculations. In particular, below the Néel temperature $T_N$ (black line), the $\operatorname{SU}(2)$-symmetry gets spontaneously broken. Hence, $\bchi$ must no longer be block-diagonal and, indeed, it acquires non-zero imaginary components in its off-diagonal blocks $\bchi_{sc}$ and $\bchi_{cs}$. In this situation, zero eigenvalues $\lambda_i=0$ in $\bchi$ result in a divergence of $\Gamma$ in the {\sl overall} (coupled) spin-charge channel.
 Then, by following the first four divergence lines from the PM phase at higher temperatures into the AF phase below $T_N$, we observe a bending to higher values of $U$ in the AF phase marked by an apparent ``kink'' at $T_N$, but without any discontinuity in accordance with the second-order nature of the phase transition. Differently, as in the PM phase, the eigenvectors associated with the zero eigenvalues thereby get a non-zero (imaginary) contribution in the spin-component $\bm{v}^s_i$ (with opposite symmetry to $\bm{v}^c_i$). This mixing is illustrated by the color scale of the corresponding (solid) divergence line, which reflects the dominant contribution to the corresponding  eigenvector $\mathbf{v}_i$, via the following expression of the respective  spin-charge imbalance:
\begin{equation}
  {\cal I}_{s-c}=\frac{\abs{\bm{v}^{sT}_i \bm{v}^s_i}-\abs{\bm{v}^{cT}_i \bm{v}^c_i}}{\abs{\bm{v}^{sT}_i \bm{v}^s_i}+\abs{\bm{v}^{cT}_i \bm{v}^c_i}},
\end{equation}
where $-1\!=\!\text{red}$ for charge, $1\!=\!\text{green}$  for spin, and $0\!=\!\text{brown}$. 
In this way, it can be readily seen how for lower $T$ and higher $U$ values, $\bm{v}^s_i$ becomes more dominant (less red, more brown), until a second eigenvalue of $\bchi$, formally associated to a spin eigenvector in the PM phase ($\bm{v}^s_i\neq0$, $\bm{v}^c_i=0$) with the same eigenvector symmetry, becomes simultaneously zero, marking an ``exceptional-point'' (EP), i.e.~a degeneracy of the matrix. Precisely at that point, the eigenvector associated with the two zero eigenvalues shows no predominant contribution, with ${\cal I}_{s-c}\!=\!0$~\cite{essl2024}. In passing the EP, one of the two eigenvalues remains zero, acquiring a dominant spin contribution in the eigenvector (solid green line), thus changing the nature of the formerly charge-dominated divergence line. This (now spin-dominated) vertex divergence line continues then towards higher $U$ values approximately following the $t^2/U$ behavior of $T_N$ in the Heisenberg regime at strong coupling.  In the same parameter regime, the second eigenvalue of the degenerate pairs at the EP, on the other hand, becomes negative, getting a charge-dominated $\mathbf{v}_i$.

At the same time, when moving from the EP towards $T=0$ and $U=0$ in the AF phase, the two eigenvalues become complex conjugate pairs with $\Re\lambda_i=0$ (and $\Im\lambda_i \neq 0$), shown as dashed brown lines. We will refer to the corresponding parameter sets as ``real-part zero-crossing'' (RZ) lines. Evidently, since their imaginary parts remain finite, they do not correspond to a divergence of $\Gamma$ \cite{vucicevic2018,essl2024}. Furthermore, we note that this condition does not exactly coincide with one of the limiting criteria for the applicability of the self-consistent perturbation theory, we reported in  \cref{eq:gamma_dmft,eq:breakdown_cond}. Hence, in general, it cannot be associated with its corresponding breakdown on a quantitative level.  Nonetheless, their occurrence marks a sign-change of the real part of the two conjugate eigenvalues from positive at higher $T$ and lower $U$ values to negative for lower $T$ and higher $U$ values, possibly relevant, as we will see, for the thermodynamic behavior of local charge (and spin) fluctuations.
Moreover, as we will discuss later on, this analysis allows us to make plausible speculations on how the perturbative breakdown may occur in the actual two-dimensional case, when the SU(2)-symmetry is restored.

In fact, as we will later investigate in \cref{sec:loc_resp}, the parameter regions where the RZs take place qualitatively correspond to the small-$U$ Slater region of the AF phase, while actual divergences occur in the large-$U$ Heisenberg regime where magnetism arises from the ordering of local moments. Indeed, even before the more detailed investigation of \cref{sec:loc_resp}, 
 this interpretation is already supported by 
a quick comparison of  RZs and vertex divergences in DMFT, with the corresponding ones in conventional MF/RPA theory [shown in panel (c) of \cref{fig:div_lines}], where only the Slater AF regime can be described. In the MF/RPA case, the local generalized susceptibility, to be compared with the DMFT one, is computed as (cf.~\cref{eq:BSE}): 
\begin{equation}
\bchi=\frac{1}{2V}\sum_\vq (\bchi_{\vq+}+\bchi_{\vq-}).
\end{equation}

As expected, since in MF/RPA the irreducible vertex $\Gamma$ cannot diverge by construction [cf. \cref{eq:gamma_rpa}],  no vertex divergence lines can be found. However, in the AF phase of the MF/RPA calculations, which has Slater nature, multiple RZ lines are found.
These RZs, originating from the point with $T=0$ and $U=0$ and extending towards the region of $T\to\infty$ and $U\to\infty$, display thus an evident similarity with those found in the Slater regime of the DMFT AF phase.

Hence, from the overall analysis of \cref{fig:div_lines} a rather definite picture emerges: Below $T_N$, one observes clear bending of the vertex divergence lines towards higher $U$ values w.r.t.~the corresponding PM phase, somewhat consistent with the expectation that the onset of a long-range order partly mitigates the correlation effects.
At the same time, the appearance of similar RZ lines in the Slater regime of the DMFT AF as well as in MF/RPA, 
indicates that, once the SU(2)-symmetry is broken, the
eigenvalues of $\bchi$ can acquire a negative real part (as they do in the strong-coupling PM regime)  even in the parameter region accessible to perturbation theory, where no divergences of $\Gamma$ occur.

\subsection{Local Charge Response in Slater and Heisenberg Regime}
\label{sec:loc_resp}

\begin{figure*}
    \centering    \includegraphics[width=\textwidth]{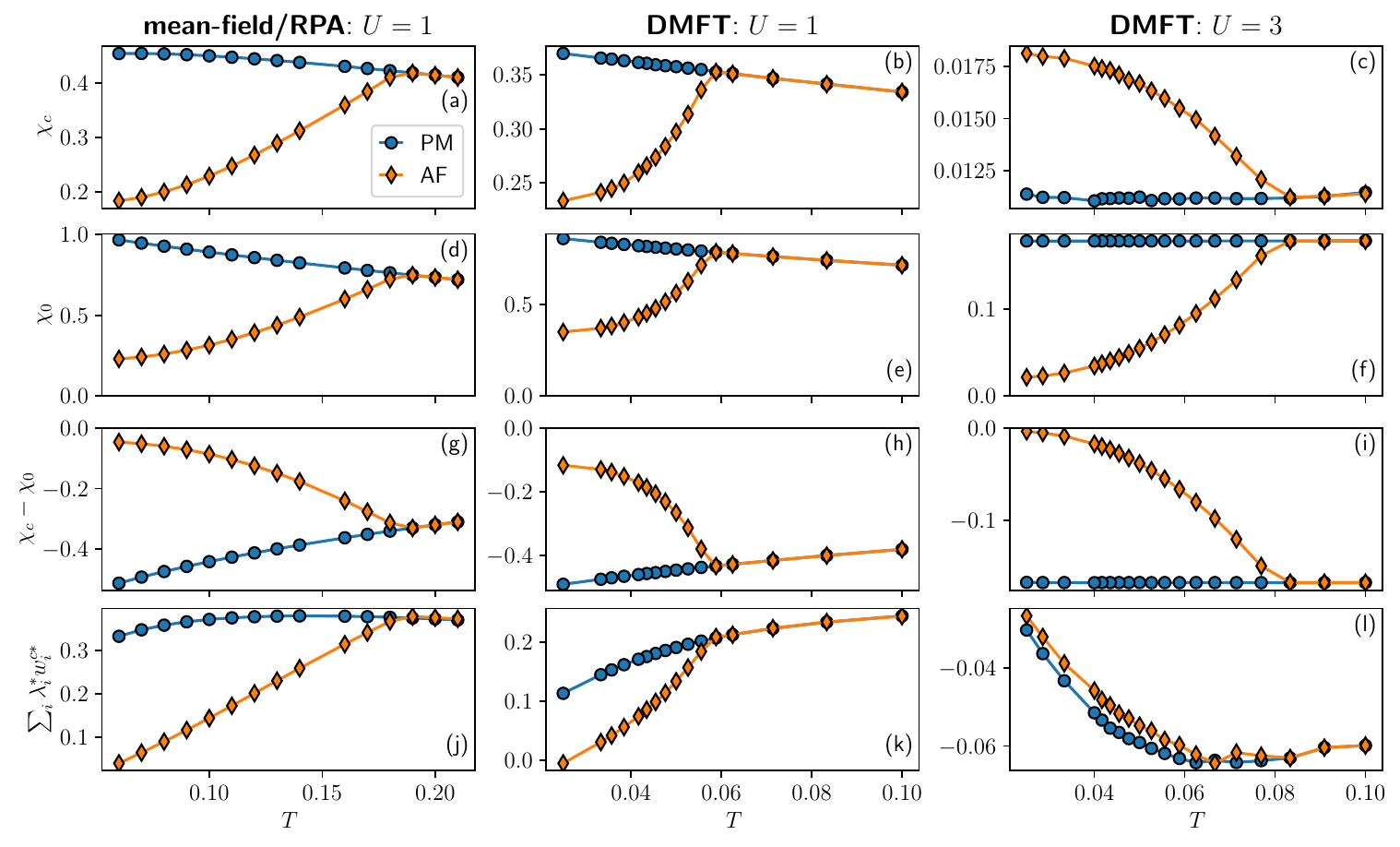}
    \caption{Comparison of the local charge response $\chi_c$ and of its different diagrammatic contributions as a function of temperature $T$  in our anti-ferromagnetic (AF, orange diamonds) and paramagnetic (PM, blue circles) calculations for weak coupling $U=1$ in mean-field/RPA (left column) and DMFT (middle column), as well as for strong coupling $U=3$ in DMFT (right column). Top row: local charge response $\chi_c$. Second-top row: bubble contribution $\chi_0$ to $\chi_c$. Second-bottom row: the difference between the bubble contribution $\chi_0$ and $\chi_c$, corresponding to the vertex corrections. Bottom row: Contributions of the eigenvalues $\lambda^*_i$ times corresponding weights $w^{c*}_i$---associated with the lowest Matsubara frequencies---to $\chi_c$.}
    \label{fig:local_ch_rsp}
\end{figure*}

To showcase the difference between the Slater and Heisenberg AF, in \cref{fig:local_ch_rsp}, we compare the \emph{local} static physical charge response $\chi_{c}=(1/\beta^2)\sum_{\nu\nu'}\bchi_c$ between the PM (blue circles) and AF phase (orange diamonds) as function of temperature $T$ for two different, representative interaction strengths.  Specifically, 
to investigate the weak coupling Slater regime, we take $U=1$ both in MF/RPA (leftmost column) and DMFT (middle column), while for the strong coupling Heisenberg regime, we fix $U=3$ in our DMFT calculations (rightmost column)\footnote{In DMFT the local static physical charge response $\chi_c$ has been obtained from a direct measurement on the imaginary time axis, in MF/RPA by analytical summation of the Matsubara frequencies. The respective contributions for $\chi_0$ and $\sum_i \lambda_i^* w_i^{c*}$ (see below) were calculated through the summation of a finite Matsubara frequency box---accounting for the high-frequency asymptotics in $\chi_0$.}. 

At $U=1$, both MF/RPA and DMFT show a qualitative similar behavior of $\chi_c(T)$ [top row, panel (a) and (b), respectively]. In particular, when AF ordering is allowed, $\chi_c$ in the AF phase appears always to decrease in comparison to the corresponding PM solutions\footnote{A similar behavior (see \cref{app:spin_response}) can be observed in the local spin response $\chi_s(T)$. In contrast to the charge response, as we will subsequently discuss, the local spin response $\chi_s(T)$ also decreases at strong coupling $U=3$ w.r.t.~the PM solution.}.

However, while in MF/RPA the qualitative behavior observed at $U=1$ persists for all interaction values of $U$ (not shown), the situation gets reversed in DMFT. Specifically, at strong coupling [$U=3$, top row, panel (c)], we observe that, although $\chi_c$ is significantly lower than for weak coupling, the AF ordering visibly enhances the \emph{local} charge response w.r.t.~the PM solution.
Such an enhancement can be indeed regarded as a characteristic hallmark of the Heisenberg AF physics, being associated with the kinetic energy gain in the AF state\cite{taranto2012,toschi2005} that arises from the onset of the AF coherence between preformed local magnetic moments.

In order to trace the origin of the different behavior of $\chi_c(T)$ between the weak-coupling Slater and strong-coupling Heisenberg AF phase at the two-particle level, we start by separating the contributions to $\chi_c(T)$ into the local ``bubble'' term $\chi_0\!=\!(1/\beta^2)\sum_{\nu\nu'}\bchi_0$, which excludes any contribution from the two-particle scattering vertex, and the ``rest'' $\chi_c\!-\!\chi_0 = \chi_{\text{vert}}$, which entails all vertex corrections. We see that in \emph{both} Slater ($U=1$) and Heisenberg regime ($U=3$), the bubble contribution $\chi_0$ is positive and becomes suppressed in the AF solution compared to the PM solution (second row in \cref{fig:local_ch_rsp}). The vertex corrections $\chi_{\text{vert}}$, instead, are negative in both regimes (also in MF/RPA) and become \emph{less negative} in the AF phase (third row in \cref{fig:local_ch_rsp}). Since bubble and vertex contributions share a similar behavior at weak and strong coupling, what actually controls the qualitative behavior of the physical response $\chi_c$ is the delicate balance between the two contributions $\chi_0$ and $\chi_{\text{vert}}$ and, in particular, whether the bubble $\chi_0$ or the vertex $\chi_{\text{vert}}$ plays the dominant role.

More insight can be gained by looking at the different eigenvalue contributions of $\bchi$ to $\chi_c$:
\begin{equation}    
    \chi_c = \sum_i \lambda_i w^c_i,
\end{equation}
where we introduced the weights of the eigenvalues to the charge sector 
\begin{equation}
    w^c_i = \left(\frac{1}{\beta}\sum_{\nu} \bm{v}^c_i\right)^2.
\end{equation}
Before proceeding an important observation should be made: For large values of the Matsubara frequencies $\nu,\nu'\to \infty$, the vertex corrections to the generalized susceptibility decay faster than the corresponding bubble term. Hence, $\bchi_c$ becomes asymptotically $\bchi_0$~\cite{rohringer2012}, i.e.~a diagonal matrix $\propto \delta_{\nu\nu'}$. Thus, for these frequency values, we find corresponding $\bchi$ eigenvalues with eigenvectors localized at high frequencies and essentially indistinguishable from those of  $\bchi_0$. Conversely, the contribution of the vertex corrections enters the eigenvalues, whose eigenvectors have a domain concentrated at low frequencies. Evidently, for those, both the vertex and $\bchi_0$ will contribute. Hence, they are the ones responsible for differentiating the Slater and Heisenberg physics. In the bottom row of \cref{fig:local_ch_rsp} we compare, between the PM and AF phase, the $\chi_c$ contributions of the eigenvalues $\lambda^*_i$ of $\bchi$ associated to eigenvectors with the lowest Matsubara frequency domains. In particular, the eigenvalues $\lambda^*_i$ of the eigenvectors $\bm{v}^{c*}_i$, that have a maximum absolute value at the lowest five Matsubara frequencies 
 \begin{equation}
    -5 \leq \underset{\mathrm{n}(\nu)}{\arg\max}\abs{\bm{v}^{c*}_i}< 5
    \label{eq:wlow}
 \end{equation}
have been considered.
At weak coupling $U=1$ (in MF/RPA and DMFT), the respective contributions $\sum_i \lambda^*_i w^{c*}_i$ show a qualitative similarity to the behavior of $\chi_0$: They are positive in the PM phase and are getting suppressed in the AF phase. At the same time, in DMFT at strong coupling, $\sum_i \lambda^*_i w^{c*}_i$ resembles \emph{qualitatively} the vertex $\chi_{\text{vert}}$: It is \emph{negative} in the PM phase, but becomes {\sl less} negative in the AF phase, contributing to the observed (relative) enhancement of $ \chi_c(T)$ below $T_N$. While the quantitative behavior of $\chi_c(T)$ cannot be rigorously reproduced without including a higher number of eigenvectors in \cref{eq:wlow}, our result demonstrates how this (low frequency) quantity already well captures the physical change occurring between the Slater and the Heisenberg AF, on a qualitative level. This observation paves the way for a thorough analysis of the eigenvalue spectrum of $\bchi$ in the different physical regimes.

\subsection{Eigenvalues of the Local Generalized Susceptibility Matrix}

\begin{figure}[th]
    \centering
    \includegraphics[width=\textwidth]{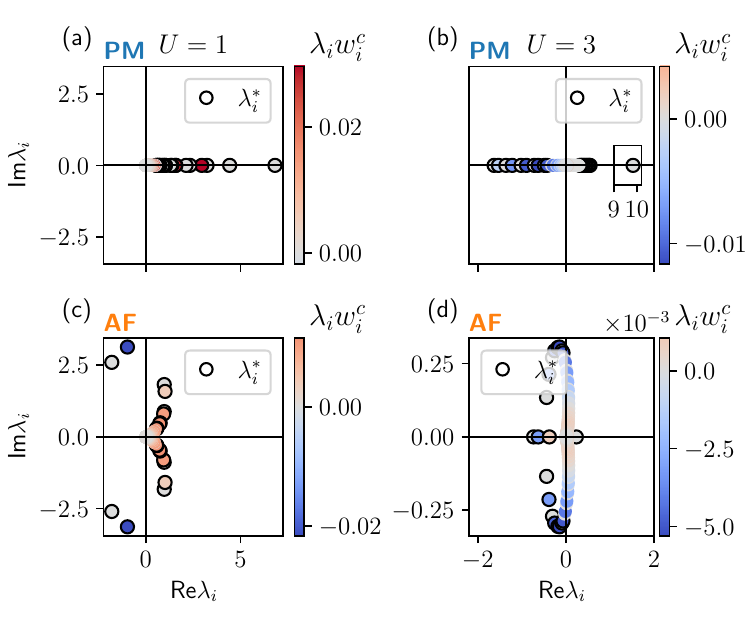}
    \caption{Eigenvalues of the local generalized susceptibility matrix $\bchi$ computed by means of DMFT, in the paramagnetic (PM, top row) and anti-ferromagnetic (AF, bottom row) phase for weak $U=1$ (left column) and strong $U=3$ (right column) coupling at $T=1/30$. The eigenvalues $\lambda^*_i$, whose eigenvectors are associated with the lowest Matsubara frequencies,  have been marked by black circles. The coloring shows the respective contribution to the local charge response $\chi_c$.}
    \label{fig:val_compare}
\end{figure}

In order to trace the hallmarks of the Slater and Heisenberg physics in the eigenspectrum of $\bchi$, in \cref{fig:val_compare} we compare the eigenvalues of the PM (top row) and AF (bottom row) solution between weak ($U=1$, left column) and strong coupling ($U=3$, right column) for a representative temperature ($T=1/30$). The color of the eigenvalues indicates the individual contributions (multiplied by their respective weights $w^c_i$) to the local charge response while the specific eigenvalues $\lambda^*_i$ ---used for the analysis in the bottom row of \cref{fig:local_ch_rsp}--- have been marked by black circles. Here, we see that the qualitatively different behavior of the contributions of $\lambda^*_i$ between weak and strong coupling is already present at the level of the individual eigenvalues. 

At weak coupling $\bchi$ (or $\bchi_c$) has only positive eigenvalues in the PM phase [panel (a), the vertex divergence lines have not been crossed], and the contributions of $\lambda^*_i$ are positive. In the AF phase [panel (c)], we see complex conjugate pairs, with a negative real part (stemming from the crossing of the RZ lines), which suppress the local charge response. 

Conversely, at strong coupling for $U=3$, in the PM [panel (b)] many vertex divergence lines have been already crossed, and hence, many negative eigenvalues are present. These lead to negative $\lambda^*_i$ contributions, which suppress $\chi_c$. In the AF phase [panel (d)], as a result of the spontaneous breaking of the SU(2)-symmetry, some of the negative real eigenvalues become complex conjugate pairs. However, all of them get a {\sl significantly less negative} real part w.r.t.~their PM counterpart, leading to the observed enhancement of the $\lambda^*_i$ contributions below $T_N$.

These considerations about the results shown in \cref{fig:val_compare} clarify the role played by the negative eigenvalues of $\bchi$  in driving the physical behavior from the Slater to the Heisenberg regime. 
In particular, the appearance of negative eigenvalues of $\bchi$ (or, more precisely, of  $\bchi_c$) in the PM phase appears to be the key ingredient for triggering an emerging AF order with Heisenberg nature: The negative real eigenvalues of the PM phase induce~\cite{gunnarsson2016,gunnarsson2017,chalupa2021,adler2024} the strong suppression of on-site charge fluctuations typical of a bad-metal/Mott insulating regime, while the breaking of the SU(2)-symmetry below $T_N$, progressively reduce their negative components, featuring the slight rebound of $\chi_c(T)$ characteristic of the Heisenberg-AF regime\footnote{The contrary conclusion cannot be made. The atomic limit of the Hubbard model has negative eigenvalues in $\bchi_c$, but shows no enhancement of the local charge response $\chi_c$, when explicitly breaking the SU(2)-symmetry by turning on an external magnetic field (see \cref{app:al_charge_resp}).}. This conclusion can be further supported by analytical considerations in the context of the Bethe lattice, see \cref{app:bethe_expansion} for details.

\begin{figure}[ht]
    \centering
    \includegraphics[width=\textwidth]{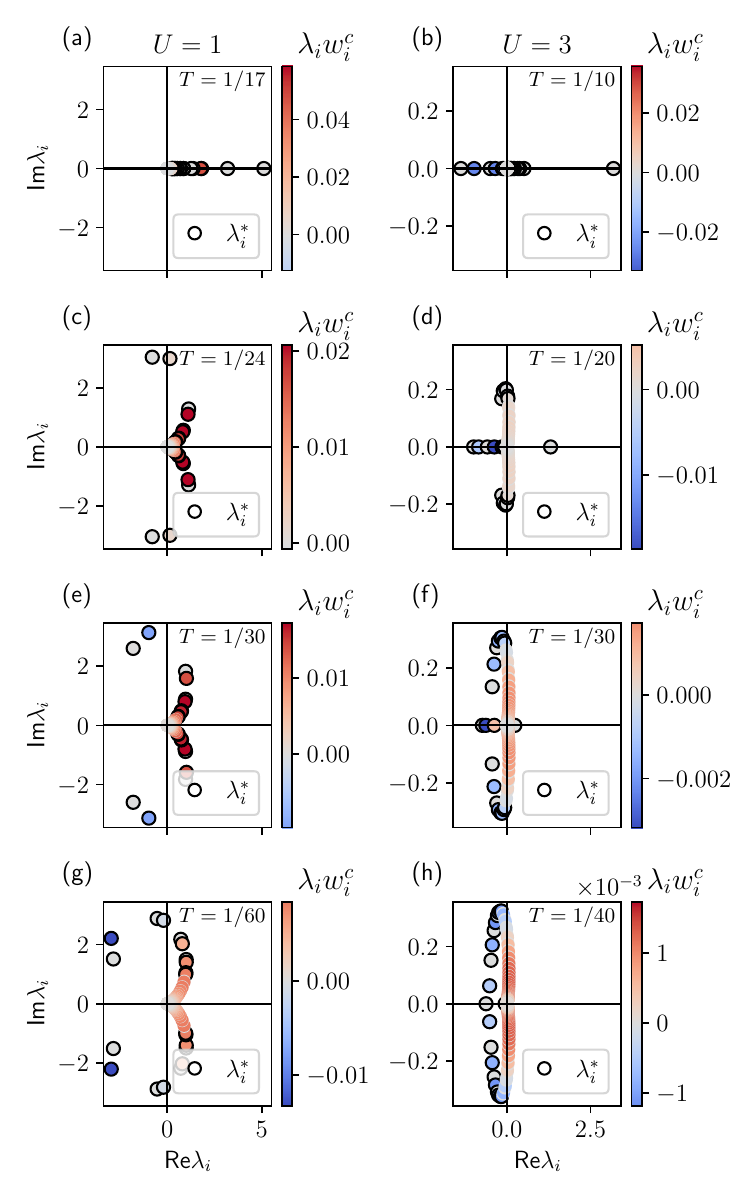}
    \caption{Similar to \cref{fig:val_compare}: Calculated DMFT eigenvalues of the local generalized susceptibility matrix $\bchi$  for weak $U=1$ (left column) and strong $U=3$ coupling (right column) for various temperatures above (first row) and below (second-to-fourth row) the corresponding $T_N$.
    }
    \label{fig:val_temperature}
\end{figure}

We complete our analysis of the local generalized charge susceptibility $\bchi_c$, by taking a look at the different eigenvalue $\lambda_i$ evolutions in temperature $T$ between weak ($U=1$, left column) and strong coupling ($U=3$, right column) when entering the AF phase in \cref{fig:val_temperature}. At weak coupling, lowering $T$ (left column, from top to bottom) and entering the AF phase [panels (c) to (g)] leads to the development of complex conjugate pairs of eigenvalues, which then appear to be disposed in a sort of ``fountain''-like structure in the representation of \cref{fig:val_temperature}. By crossing several RZ lines, more and more of them start to develop a negative real part.

At strong coupling, the behavior is more complex: Starting from negative values in the PM phase at high temperature [$T=1/10$, panel (b)] and reducing the temperature (right column, top to bottom) leads, below $T_N$ [panels (d) to (h)], to the emergence of similar ``fountain''-like structure as at weak coupling, albeit with a relevant difference: Not all eigenvalues of $\bchi$ become immediately complex conjugate pairs at the breaking of the SU(2)-symmetry. In particular, the most negative eigenvalues of the PM phase stay, at first, real also below $T_N$, but become progressively less negative. At the same time, also the most positive real eigenvalues of $\bchi$ with a predominant spin contribution of their associated eigenvectors remain real below $T_N$. Their value, however, gets gradually suppressed, until they cross zero at spin-dominated vertex divergence lines [greenish lines in panel (b) of  \cref{fig:div_lines}].
By further reducing the temperature, these eigenvalues become negative\footnote{We note that these negative real eigenvalues with a predominant spin contribution, when real and antisymmetric in $v^s_i$, and imaginary and symmetric in $v^c_i$, actually also do contribute enhancing to the charge response $\lambda_iw^c_i$.} until they eventually meet the previously discussed negative real eigenvalues. Their merging gives rise to the formation of complex conjugate pairs.  The largest real eigenvalue, colored in gray provides, indeed, a good example of this general behavior in the right column of \cref{fig:val_temperature}. It is also interesting to emphasize that this specific eigenvalue mostly contributes to the enhanced Curie-like static magnetic response of the PM bad-metallic/Mott phase~\cite{springer2020}. Hence, it encodes, within the spin scattering channel, a significant aspect of the local moment physics, which underlies the onset of a Heisenberg AF instability. Quite remarkably, it is precisely the vanishing of this eigenvalue in the strong-coupling AF phase that determines the appearance of the first vertex divergence line with a predominant spin nature in the Heisenberg regime of the AF phase.

\section{BSE in the AF Phase: An Eigenvalue Perspective}
\label{AFbse}

In Refs.~\cite{reitner2020,delre2021fluctuations,reitner2024,kowalski2024,loon2024}, in the context of DMFT, the eigenvalues of the static local generalized susceptibility $\bchi$ have been successfully linked to the values of the uniform ($\fq=0$) and/or staggered [$\fq= (\pi, \pi, \cdots)$] linear response of the lattice system considered.
In particular, it was demonstrated that the largest positive and lowest negative eigenvalues of the local $\bchi$ could directly trigger actual thermodynamic instabilities. It is thus natural to ask, to what extent these findings generalize to the AF phase of the bipartite lattice. To this aim, we start by recalling that, within DMFT, the BSE in the PM phase---for both, spin and charge susceptibilities---takes the following form~\cite{georges1996,reitner2020,delre2021fluctuations} [cf. with \cref{eq:dmft_bse}]
\begin{equation}
\label{eq:bse_pm}
    \bchi_{\fq c/s} = \left[\bchi^{-1}_{c/s} +\Delta\bchi^{-1}_{0\fq} \right]^{-1}.
\end{equation}
 Here, all quantities are defined on the single atomic basis, hence $\fq$ runs over the full BZ, and we defined $\Delta\bchi^{-1}_{0\fq} = \bchi^{-1}_{0\fq}-\bchi^{-1}_{0}$---analog to the bipartite lattice. For the Bethe-lattice case (in the infinite coordination limit, with  rescaled nearest neighboring hopping $t$), \cref{eq:bse_pm} becomes particularly intriguing, as its explicit expression connects \cite{georges1996,reitner2020} directly the eigenvalues $\lambda^{c/s}_i$ of the local quantity $\bchi_{c/s}$ with the eigenvalues $\lambda^{\fq c/s}_i$ of $\bchi_\fq$ \cite{georges1996,reitner2020,delre2021fluctuations,loon2022,kowalski2024}:
 \begin{align}
    \label{eq:bethe_chi_q0}
    \bchi_{\fq=(0,0,\dots) c/s} =& \left[\bchi^{-1}_{c/s} +\frac{t^2}{\beta}\mathbbm{1} \right]^{-1},\\
    \label{eq:bethe_chi_qpi}
    \bchi_{\fq=(\pi,\pi,\dots) c/s} =& \left[\bchi^{-1}_{c/s} -\frac{t^2}{\beta}\mathbbm{1} \right]^{-1},\\
    \label{eq:bethe_lambda_q0}
    \lambda_i^{\fq=(0,0,\dots) c/s} =& \left( \frac{1}{\lambda_i^{c/s}} +\frac{t^2}{\beta}\right)^{-1},\\
      \label{eq:bethe_lambda_qpi}
    \lambda_i^{\fq=(\pi,\pi,\dots) c/s} =& \left( \frac{1}{\lambda_i^{c/s}} -\frac{t^2}{\beta}\right)^{-1}.
\end{align}
 An eigenvalue $1/\lambda_i^{c/s}$ reaching $\pm t^2/\beta$ thus triggers a divergence of $\lambda_i^{\fq, c/s}$ and, if the corresponding weight $w^{c/s}_i$ is finite, also a divergence of the corresponding physical response 
\begin{equation}
    \chi_{\fq c/s}=\frac{1}{\beta^2}\sum_{\nu\nu'}\bchi_{\fq c/s} = \sum_i \lambda_i^{\fq c/s} w^{c/s}_i.
\end{equation}
 Thereby in the PM phase $1/\lambda_i^{s}\to t^2/\beta$  is responsible for the divergence of the AF spin response at the AF phase transition~\cite{delre2021fluctuations} and $1/\lambda_i^{c}\to -t^2/\beta$ for the divergence of the derivative of the double-occupancy $\partial D/\partial U$ Ref.~\cite{kowalski2024}, as well as for the divergence of the uniform charge response at the critical point of the MIT away from \ph-symmetry~\cite{reitner2020,loon2022,reitner2024}. For the case of the square lattice of interest in this study, \cref{eq:bethe_chi_q0,eq:bethe_chi_qpi,eq:bethe_lambda_q0,eq:bethe_lambda_qpi} have been numerically found to hold approximately \cite{reitner2020,moghadas2025} by replacing $t\to t_{\text{eff}}(\mu,\beta,U)$ with an effective constant that depends on the system parameters, with a good agreement to the Bethe lattice expression in the intermediate-to-strong coupling regime.

\begin{figure}[ht]
    \includegraphics[width=\textwidth]{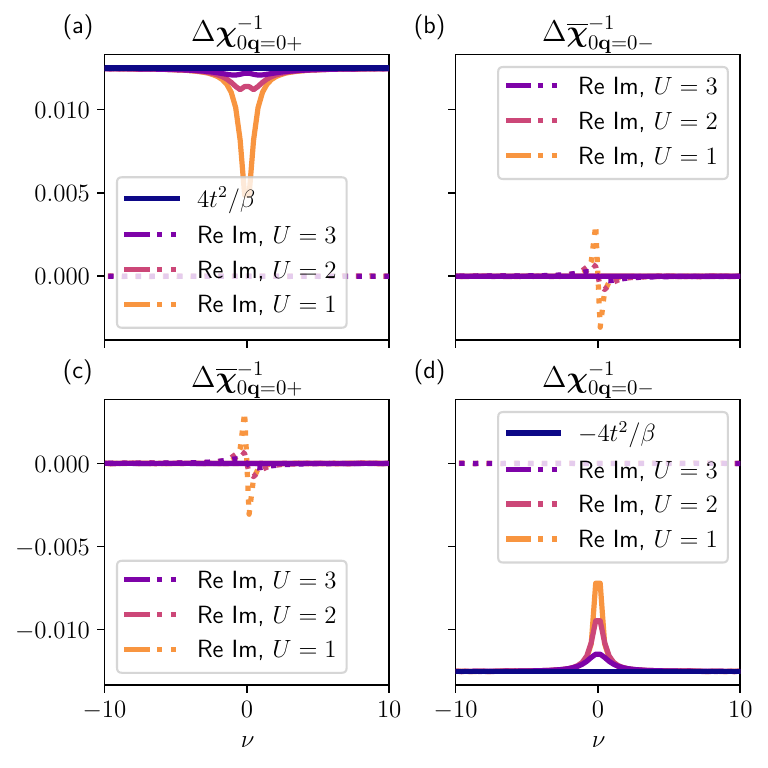}
    \caption[]{Difference of the inverse uniform and local bubble $\Delta\bchi^{-1}_{0\vq=0}$ as function of $\nu=\nu'$ in the AF phase of the square lattice Hubbard model solved in DMFT for different values of the Hubbard interaction from weak $U=1$ to strong $U=3$ coupling at temperature $T=1/20$.}
    \label{fig:t_eff}
\end{figure}

We now generalize these relations to the DMFT solution for the AF phase in a bipartite lattice. As we show in \cref{app:proof_bethe}, in the Bethe lattice, \cref{eq:dmft_bse} explicitly reads
\begin{equation}
\label{eq:af_bse_bethe}
    \bchi_{\vq=0\pm} = \left[\begin{pmatrix}
        \bchi_s & \bchi_{sc}\\
        \bchi_{cs} & \bchi_c
    \end{pmatrix}^{-1}
    \pm \frac{t^2}{\beta} 
    \begin{pmatrix}
        \mathbbm{1} & 0\\
        0 & -\mathbbm{1}
    \end{pmatrix}
    \right]^{-1},
\end{equation}
featuring the direct  generalization of the PM expressions of \cref{eq:bethe_chi_q0,eq:bethe_chi_qpi}.

As a next step, the evaluation of  $\Delta\bchi^{-1}_{0\vq=0}$ for the bipartite square lattice in \cref{fig:t_eff} at different values of $U$ shows that similarly as in the PM, also in the AF-phase, the Bethe lattice expression \cref{eq:af_bse_bethe} (with the re-scaled hopping $4t^2/\beta$ in two-dimensions~\cite{moghadas2025})  becomes a good approximation at intermediate-to-strong coupling. 

Given this result, it would then be tempting to search for an analogous direct connection between the eigenvalues of the local $\bchi$ and the lattice response  $\bchi_{\vq=0\pm}$ such as the one valid in the PM phase. However, due to the $\sigma_z$-like structure of $\Delta\bchi^{-1}_{0\vq=0}$, which ultimately originates from the checkerboard pattern of the AF order on the bipartite lattice, the identification of a so direct link is no longer possible. In fact, in contrast to the case of a true multiorbital system and/or of a ferromagnetic order \cite{kowalski2024} (where instead of a Pauli matrix $\sigma_z$ structure, one simply has the identity $\mathbbm{1}$), in the AF-DMFT case, it is not possible to diagonalize simultaneously the two terms on the r.h.s.~of \cref{eq:af_bse_bethe}.

Nonetheless, at small values of the magnetization $m$, when $\bchi_{sc}$ and $\bchi_{cs}$ are small compared to $\bchi_s$ and $\bchi_c$, we can employ a different strategy:
We transform \cref{eq:af_bse_bethe} into the eigenbasis of the respective subspaces of the local spin $\bchi^{\phantom{s}}_s \bv^s_i = \lambda^s_i \bv^s_i$ and charge $\bchi^{\phantom{c}}_c \bv^c_i =\lambda^c_i \bv^c_i $ susceptibilities: 
\begin{equation}
\begin{split}
\label{eq:bse_proj}
&\begin{pmatrix}
    \bv^{s}_i & 0\\
    0 & \bv^c_i
\end{pmatrix}^T
\bchi_{\vq=0 \pm}
\begin{pmatrix}
    \bv^s_j & 0\\
    0 & \bv^c_j
\end{pmatrix}\\
&= 
\left[\begin{pmatrix}
        \lambda^s_i\delta_{ij} & \lambda^{sc}_{ij}\\
        \lambda^{sc}_{ji} &  \lambda^c_i\delta_{ij}
\end{pmatrix}^{-1} \pm \frac{1}{\beta}
\begin{pmatrix}
         t^2 & 0\\
        0 & - t^2
    \end{pmatrix}\delta_{ij}\right]^{-1},
\end{split}
\end{equation}
 where it should be noted that, while $(\bv^s_i,0)$ and $(0,\bv^c_i)$ are no eigenvectors of the general $\bchi$ matrix, they do form an orthonormal basis set.
Then, if two subspace eigenvectors $(\bv^s_i,0)$, $(0,\bv^c_i)$ span the same eigenvector space as two eigenvectors $\mathbf{v}_i$ of the total $\bchi \bm{\mv}_i = \lambda_i \bm{\mv}_i$, we have
 \begin{align}
    \label{eq:v_approx_sc}
     \bv^{sT}_i\bchi_{sc}\bv^c_j &= \lambda^{sc}_{ij} \approx  \lambda^{sc}_i \delta_{ij},\\
     \label{eq:v_approx_cs}
     \bv^{cT}_i\bchi_{cs}\bv^s_j &= \left(\bv^{sT}_j\bchi_{sc}\bv^c_i\right)^T =\lambda^{sc}_{ji}\approx  \lambda^{sc}_i \delta_{ji},
 \end{align}
 where $\lambda^{sc}_{ij}$ becomes purely imaginary at \ph-symmetry.
 
\begin{figure}[ht]
    \includegraphics[width=\textwidth]{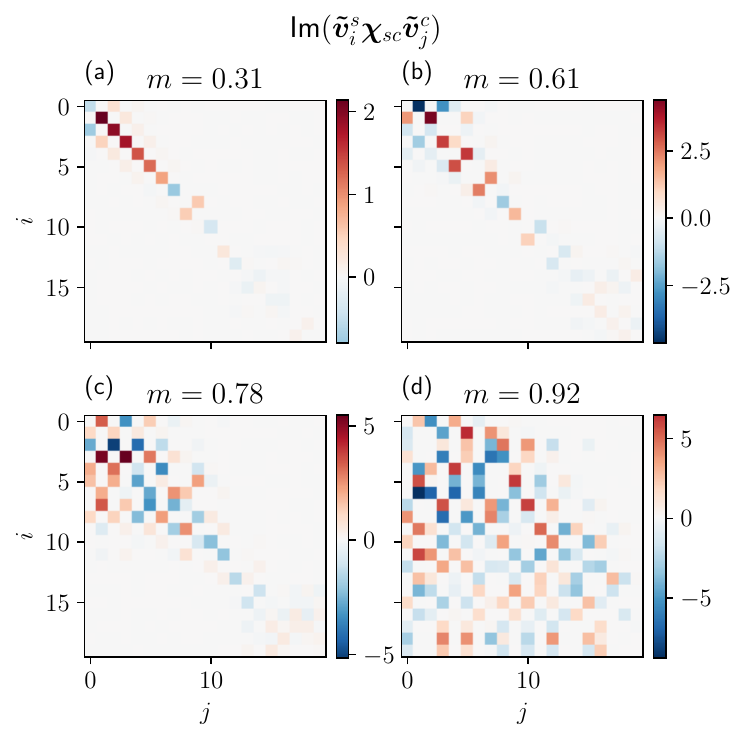}
    \caption[]{$\Im(\bv_i^s \bchi_{sc} \bv_j^c)$: imaginary part of the projection of the symmetry breaking local generalized $\bchi_{sc}$ susceptibility onto the eigenvectors $\bv^s_i,\bv^c_i$ of $\bchi_{s}$ and $\bchi_{c}$  for interaction value $U=3$ and with increasing magnetization $m$ at temperatures $T=(1/13,1/15,1/18,1/30)$. One can easily notice how the approximated relation of \cref{eq:v_approx_sc,eq:v_approx_cs} holds reasonably well for small $m$, while it deteriorates for large magnetization values.}
    \label{fig:sc_projection}
\end{figure}

 In \cref{fig:sc_projection}, we can verify, via our DMFT calculations, that this approximation actually holds for small values of the magnetization $m$  [panel (a)] and that, hence,  the BSE equation can be separated into distinct $2\times2$ matrix equations. As expected, the approximation breaks down with increasing $m$, because, then, larger clusters of subspace eigenvectors couple to each other.

After having verified the condition under which the approximation in \cref{eq:v_approx_sc,eq:v_approx_cs} holds, let us shortly discuss their consequences in the regime of its applicability.
The right hand side of \cref{eq:bse_proj} would yield:
 \begin{equation}
 \begin{split}
 \label{eq:af_2x2}
    & \frac{D_i \delta_{ij}}{(\frac{1}{\lambda^s_i}\pm \frac{t^2}{\beta} D_i)(\frac{1}{\lambda^c_i} \mp \frac{t^2}{\beta} D_i) -(\frac{\lambda^{sc}_i}{\lambda^s_i \lambda^c_i})^2 }\\ 
     &\times\begin{pmatrix}
         \frac{1}{\lambda^c_i} \mp \frac{t^2}{\beta} D_i  & \frac{\lambda^{sc}_i}{\lambda^s_i \lambda^c_i}\\
        \frac{\lambda^{sc}_i}{\lambda^s_i \lambda^c_i} & \frac{1}{\lambda^s_i}\pm \frac{t^2}{\beta} D_i
    \end{pmatrix},
\end{split}
\end{equation}
where 
\begin{equation}
    D_i= 1 -\frac{(\lambda^{sc}_i)^2}{\lambda^s_i \lambda^c_i}. 
\end{equation}
Here, we can immediately notice that in the AF phase, the hopping constant $t^2 \to t^2 D_i$  appears as if it were effectively renormalized in the BSE. At the same time, even taking into account such renormalization, one could no longer directly apply analogous instability conditions as in the PM phase, such as:
\begin{align}
\frac{1}{\lambda^{s}_i}\pm \frac{t^2}{\beta} D_i &=0,\\
\frac{1}{\lambda^{c}_i}\mp \frac{t^2}{\beta} D_i &=0,
\end{align}
  due to the finite offset $-(\lambda^{sc}_i/\lambda^s_i \lambda^c_i)^2$. On the contrary, quite interestingly, an instability for $\lambda^{sc}_i\!\neq\!0$ can now only appear simultaneously  both in the spin and in the charge component of the BSE, when the following relation:
\begin{equation}
    (\frac{1}{\lambda^s_i}\pm \frac{t^2}{\beta} D_i)(\frac{1}{\lambda^c_i} \mp \frac{t^2}{\beta} D_i) -(\frac{\lambda^{sc}_i}{\lambda^s_i \lambda^c_i})^2 =0
\end{equation}
holds.
We note here that when approaching the phase transition, the magnetic order reduces sufficiently fast to yield $\lambda^{sc}_i\!\to\!0$ and $D_i\sim \mathcal{O}[(\lambda^{sc}_i)^2]\!\to \!1$, so that the AF instability condition 
\begin{equation}
\label{eq:af_cond}
    \frac{1}{\lambda^{s}_i} - \frac{t^2}{\beta} = 0
\end{equation}
coincides in the AF and PM phase. This implies that, when approaching the magnetic transition in the AF phase for $\lambda^{sc}_i\!\to\!0$,  where the AF condition in \cref{eq:af_cond} is fulfilled,  also the off-diagonal components in \cref{eq:af_2x2} can be expected to diverge. In contrast, the charge component, for $\lambda^c_i\neq-\beta/t^2$, remains finite.
We want to emphasize here, that while these considerations rigorously hold only in the limit of small $m$, the main effects of the SU(2) symmetry-breaking in the BSE for the AF phase (i.e., (i) a sizable renormalization of $t^2$ term and (ii) a global instability condition that differs from the PM ones) can be expected to have a general validity.

At particle-hole symmetry, however,  due to the coupled symmetry of the eigenvectors for $\bm{\chi}_{\vq=0\pm}$ (either symmetric in spin $\bm{v}_i^s$ and antisymmetric in charge $\bm{v}_i^c$, or vice versa), the instability of one eigenvalue of $\bm{\chi}_{\vq=0\pm}$ can only occur for the charge or spin sector:  its effects will be, thus,  \emph{not} transmitted to the other physical response functions in \cref{eq:dndmu,eq:dndh,eq:dmdmu,eq:dmdh}, since the antisymmetric components vanish in the Matsubara sum. For finite doping instead, where particle-hole symmetry is violated, formerly antisymmetric eigenvectors acquire a symmetric part~\cite{reitner2024}, and do contribute to the physical response functions~\cite{reitner2020,kowalski2024}. There, a simultaneous instability in both the spin and charge response can be expected. Possibly, this occurs due to a similar mechanism as in Ref.~\cite{reitner2020}.

\section{\label{Conclusion}Conclusions}

In this work, we have presented a thorough investigation of the divergences of the irreducible vertex functions in the presence of a spontaneously broken SU(2)-symmetry. This way, we lifted one of the limitations of literature studies \cite{schafer2013,janis2014,kozik2015, stan2015,rossi2015,rossi2016,gunnarsson2016,schafer2016,tarantino2018,vucicevic2018,chalupa2018,springer2020,reitner2020,chalupa2021,mazitov2022localmomentA,mazitov2022localmomentB,pelz2023,adler2024,reitner2024,kowalski2024} on the breakdown of the many-electron self-consistent perturbation theory, which were hitherto restricted to paramagnetic phases. 

In particular, we performed DMFT calculations on the two-particle level in the AF phase of the half-filled (particle-hole symmetric) Hubbard model, from weak to intermediate-to-strong coupling, determining the location and the nature of the divergences of the irreducible vertex functions in the $T\! - \! U$ phase- diagram of the problem.

The main findings of our analysis can be summarized as follows:
(i) The onset of the AF long-range order tends to mitigate the occurrence of divergences in the irreducible vertex functions: For decreasing temperatures and for increasing values of the staggered magnetization, their location gets progressively shifted towards larger values of $U$ w.r.t.~the underlying paramagnetic phase.

(ii) In the AF phase, due to the spontaneous breaking of the SU(2)-symmetry,  the charge and the longitudinal sectors of the Bethe Salpeter equation for the generalized susceptibility $\bchi$ become coupled  \cite{bickers2004,delre2021su2,essl2024}. Hence,  below $T_N$, the nature of the corresponding vertex divergencies acquires a mixed character. In particular, our analysis of the eigenvectors associated with vanishing eigenvalues of $\bchi$ (and hence to the vertex divergencies) indicates that right below $T_N$, the vertex divergencies display a predominating charge character (prevalently red colored lines in Fig.~\ref{fig:div_lines}), similar to the one they had in the PM phase. However,  when reducing the temperature further, this character abruptly changes into a dominating longitudinal spin one (prevalently green colored lines in Fig.~\ref{fig:div_lines}) after the occurrence of an exceptional point.

Quite significantly, the above-mentioned exceptional point is found when the largest eigenvalue of $\bchi$ in the spin sector of the PM solution, responsible \cite{springer2020} for the characteristic Curie behavior of the physical magnetic response in the Mott phase, is damped down to zero for increasing value of the magnetization. This provides a direct link between observed vertex divergencies in the AF phase and the physics of the local moment formation.

(iii) From any exceptional point, a RZ line (where the real part of a pair of complex conjugate eigenvalues of $\bchi$ is zero, albeit with a finite imaginary part) departs toward the origin of the $T\!-\!U$-phase-diagram. 

As a consequence of (i)-(iii), the AF-phase-diagram of the Hubbard model in DMFT gets roughly split into two parts w.r.t.~the irreducible vertex properties: At weak-coupling, below $T_N$, only RZ lines are present, qualitatively similar to those we found in the (static) RPA calculations.  Actual vertex divergences, systematically shifted to larger $U$ values w.r.t.~the PM phase, are found, instead, in the intermediate-to-strong coupling regime.

The  qualitative difference in the two-particle properties gets directly reflected in the different nature of the AF phases in the two regimes (Slater at weak-coupling vs. Heisenberg at strong-coupling), as 
demonstrated by our comparative analysis of the contributions of the eigenvalues/eigenvectors of $\bchi$ to the behavior of the respective local static charge response.

While the DMFT expressions relating the eigenspectrum properties of the local $\bchi$ to the physical response of the lattice system become more complicated than in the SU(2)-symmetric case \cite{georges1996, vanloon2020,reitner2020,reitner2024,kowalski2024}, at temperatures slightly below $T_N$ (i.e., for small values of the staggered magnetization $m$)  physically transparent approximations could be derived and successfully tested. These might be useful, e.g., to investigate the occurrence of phase-separation instabilities in the AF-phase out-of-half filling \cite{arzhnikov2012,lenihan2022,rampon2025}. This latter topic has indeed a long history related to the proposal that phase separation can be a crucial driving force of the cuprate phase diagram as a function of doping, leading to both charge-density-wave ordering and to superconductivity itself~\cite{Castellani1995}. In this regard, an analysis based on two-particle properties can shed light on the role of magnetism and of the reduced dimensionality~\cite{sorella2022} in driving the phase separation instability and in its relation with superconductivity.

 \section{\label{Outlook}Outlook}
 
The results obtained in our study have several implications, both for future algorithmic developments and for a more fundamental understanding of the many-electron theory and the effect of symmetry breaking on its properties.
 
From an algorithmic perspective, the distinct properties of two-particle DMFT vertex functions in the weak- and strong-coupling regimes---encoding the differences between Slater and Heisenberg antiferromagnets---offer valuable insights for generalizing diagrammatic extensions of DMFT \cite{rohringer2018}, such as D$\Gamma$A \cite{toschi2007} and DMF$^2$RG \cite{taranto2014,wentzell2015}, to study AF-ordered phases \cite{delre2021su2,delre2025}.

In particular, our findings indicate the necessity to choose, as local input for diagrammatic resummations beyond DMFT,  local two-particle vertex functions {\sl directly} calculated in the corresponding AF-ordered phase.  In fact, if one stabilized the AF instabilities only through diagrammatic resummations built on a PM DMFT input, essential aspects of the non-perturbative AF-physics (such as the crossover from Slater to Heisenberg \cite{sangiovanni2006,taranto2012}) would be completely absent in the theoretical description.

From a more physical viewpoint,  RZ lines are present in both the AF-solution of RPA as well as in the weak-coupling/Slater regime of DMFT, displaying an exponential behavior down to $T\!=\!U=\!0$. This provides---as we will detail in the following---intriguing hints on possible non-perturbative features of the putative exact solution of two-dimensional systems \cite{qin2022}. There, due to the Mermin-Wagner theorem \cite{mermin1966}, the onset of an AF-order will be confined to the ground state, while, at finite $T$, the ordered phase will be replaced by a regime of quasi-long-range \cite{baier2004,jenkins2022} (i.e., with an exponentially growing magnetic correlation length \cite{dare1996,tremblay2011,schafer2015,schafer2021}) AF fluctuations. In this situation, the ``formally''  preserved $SU(2)$-symmetry of our (particle-holy symmetric) problem will render $\bchi$ a real symmetric matrix, block diagonal in the charge and in the spin sector, with a purely real eigenvalue spectrum.
As a consequence the expected suppression of the charge response by strong (long-range) AF fluctuations~\cite{gull2008} can no longer be realized through the same mechanism as in the  Slater regime described by the ``perturbative'' RPA. There, the SU(2)-symmetry breaking, allowing for eigenvalues with non-zero imaginary part, leads to  the ``fountain structure'' visible in Figs.~\ref{fig:val_compare}-\ref{fig:val_temperature}, which suppresses the charge fluctuations.
In the exact 2D solution, which fulfills the Mermin-Wagner theorem,  the lack of long-range order will suppress the imaginary parts of the eigenvalues. In that case, the progressive reduction of on-site charge fluctuations will plausibly occur through the true vanishing of purely real eigenvalues.

 Hence, the RZ lines found in the Slater regime would transform into actual divergence lines. In this respect, it would then be plausible to expect, in the exact solution of the two-dimensional Hubbard model, an accumulation of vertex divergence lines towards $T\!= \! U \!= \! 0$, with a qualitatively similar (exponential) shape as the RZ lines in our AF phase diagram of the DMFT solution.
In fact, such expectation would also be consistent with the trend of the very few existing cluster-DMFT \cite{maier2005} studies \cite{gunnarsson2016,vucicevic2018} of vertex divergencies in the PM phase of the two-dimensional Hubbard model, which showed an evident shift of the divergence lines towards lower $U$ values w.r.t.~the corresponding DMFT calculations.

Evidently, such speculations call for future, systematic cluster-DMFT investigations of the 2D Hubbard model on the two-particle level and the application of the more recently introduced methods~\cite{bonetti2022,goremykin2024} that restore the Mermin-Wagner theorem, starting from a symmetry-broken (dynamical) mean-field solution.
In particular, the verification of such a scenario would provide a clear indication, in a rigorous theoretical framework, for a complete vanishing of the convergence radius of the self-consistent perturbation theory for the (particle-hole symmetric) two-dimensional Hubbard model in the zero temperature limit.

\begin{acknowledgments}
\noindent 
 We thank P.~Chalupa-Gantner, H.~Eßl, E.~Moghadas, P.~Oberleitner, S.~Rohshap, G.~Sangiovanni, and D.~Springer for helpful comments and discussion. 
M.C. acknowledges financial support from the
National Recovery and Resilience Plan (NRRP) MUR Project  No. CN00000013-ICSC and PE0000023-NQSTI and by  MUR via PRIN 2020 (Prot. 2020JLZ52N 002) programs, PRIN 2022 (Prot. 20228YCYY7). 
 M.R. acknowledges support as a recipient of a DOC fellowship of the Austrian Academy of Sciences.
 This research was also funded by the Austrian Science Fund (FWF) via [doi:10.55776/I5487] (M.R.) and [doi:10.55776/I5868], Project No. P1 of the research unit FOR 5249 of the German research foundation (DFG) (A.T.). For open access purposes, the corresponding author (M.R.) has applied a CC BY public copyright license to any author accepted manuscript version arising from this submission.
 Calculations have been performed on the Vienna Scientific Cluster (VSC). 
\end{acknowledgments}
\FloatBarrier

\appendix
\section{Data Availability}
\label{app:data}
 A data set containing all numerical data and plot scripts used to generate the figures of this publication is publicly available at \cite{data}.

\section{Symmetry of the Static Generalized Susceptibility}
\label{app:chi_symm}
At \ph-symmetry (half-filling) the generalized susceptibility in \cref{eq:chi} is invariant under exchanging creation and annihilation operators with opposite spin $\create{\alpha\sigma} \leftrightarrow \annihil{\alpha-\sigma}$ also in the AF phase \cite{essl2024}. Hence, any summation of opposite spin components $\bchi_{\sigma\sigma'}+\bchi_{-\sigma-\sigma'}$ becomes a real and bisymmetric matrix in Matsubara frequencies \cite{rohringer2012,springer2020,essl2024}. Therefore, $\bchi_{s}$, $\bchi_{c}$ become real and bisymmetric matrices, while $\bchi_{sc}$, $\bchi_{cs}$, composed of the opposite spin component differences, become purely imaginary centro-Hermitian matrices \cite{lee1980,reitner2024} at \ph-symmetry (non-zero in the AF phase). Away from \ph-symmetry all, $\bchi_{s}$, $\bchi_{c}$, $\bchi_{sc}$, and $\bchi_{cs}$, become complex centro-Hermitian matrices.

Further, by exchanging $\nu \leftrightarrow \nu'$ (for $\omega=0$), corresponding to a matrix transpose, the spin components in \cref{eq:chi} are exchanged, leading to $\bchi_{ \sigma\sigma'}^T = \bchi_{ \sigma'\sigma}$. From this and the considerations above we conclude that $\bchi$ is (i) a symmetric $\bchi^T=\bchi$ and (ii) a $\kappa$-real matrix \cite{hill1992,reitner2024,essl2024}:
\begin{equation}
    \bm{\Pi}\bchi\bm{\Pi} = \begin{pmatrix}
        \bm{J}\bchi_s\bm{J} &  \bm{J}\bchi_{sc}\bm{J}\\
        \bm{J}\bchi_{cs}\bm{J} & \bm{J}\bchi_c\bm{J}
    \end{pmatrix} = \begin{pmatrix}
        \bchi^*_s &  \bchi^*_{sc}\\
        \bchi^*_{cs} & \bchi^*_c
    \end{pmatrix} =\bchi^*,
\end{equation}
at \ph-symmetry
\begin{equation}
    \bm{\Pi}\bchi\bm{\Pi} = \begin{pmatrix}
        \phantom{-}\bchi_s &  -\bchi_{sc}\\
        -\bchi_{cs} & \phantom{-}\bchi_c
    \end{pmatrix} =\bchi^*,
\end{equation}
with 
\begin{equation}
   \bm{\Pi} = \begin{pmatrix} \bm{J} &0\\0&\bm{J}\end{pmatrix}
\end{equation}
being a permutation matrix and $\bm{J}\coloneqq \delta_{\nu(-\nu')}$.

\section{Symmetry of the Eigenvectors at \ph-symmetry}
\label{app:v_symm}
At \ph-symmetry the generalized susceptibility in the longitudinal channel 
\begin{equation}
    \bchi = \begin{pmatrix}
        \bchi_s & \bchi_{sc}\\
        \bchi_{cs} & \bchi_c
    \end{pmatrix}
\end{equation}
consists of two bisymmetric real sub-matrices $\bchi_s\!=\!\bm{J}\bchi_s\bm{J}$, $\bchi_c\!=\!\bm{J}\bchi_c\bm{J}$ and of two pure imaginary centro-Hermitian sub-matrices $\bchi_{sc}=-\bm{J}\bchi_{sc}\bm{J}$, $\bchi_{cs}=-\bm{J}\bchi_{cs}\bm{J}$, with $\bchi_{sc}=\bchi_{cs}^T$, where $\bm{J}\coloneqq\delta_{\nu(-\nu')}$. Hence, we can further split the sub-matrices into the following $2\times2$ blocks~\cite{essl2024}:
\begin{equation}
    \bchi = 
    \begin{pmatrix}
    \bm{A} & \bm{JBJ} & \bm{E}   & -\bm{JFJ} \\
    \bm{B} & \bm{JAJ} & \bm{F}   & -\bm{JEJ} \\
    \bm{E} & \bm{F}   & \bm{C}   & \bm{JDJ} \\
    \bm{-JFJ} & \bm{-JEJ}   & \bm{D}   & \bm{JCJ} \\
    \end{pmatrix}.
\end{equation}
Let us formally split the eigenvectors $\mathbf{v}_i$ of $\bchi$, $\bchi\mathbf{v}_i=\lambda_i \mathbf{v}_i$, into symmetric and anti-symmetric parts
\begin{equation}
    \mathbf{v}_i = 
    \begin{pmatrix}
        \bm{s}^s_i\\
        \bm{Js}^s_i\\
        \bm{s}^c_i\\
        \bm{Js}^c_i\\
    \end{pmatrix}
    +
     \begin{pmatrix}
        \bm{a}^s_i\\
        -\bm{Ja}^s_i\\
        \bm{a}^c_i\\
        -\bm{Ja}^c_i\\
    \end{pmatrix}
\end{equation}
leading to
\begin{equation}
\begin{split}
    \bchi\mathbf{v}_i= &
    \begin{pmatrix}
    \phantom{-\bm{J}[}(\bm{A}+\bm{JB})\bm{s}^s_i + (\bm{E}+\bm{JF})\bm{a}^c_i \phantom{]}\\
    \phantom{-}\bm{J}\left[(\bm{A}+\bm{JB})\bm{s}^s_i + (\bm{E}+\bm{JF})\bm{a}^c_i \right]\\
     \phantom{-\bm{J}[}(\bm{C}-\bm{JD})\bm{a}^c_i + (\bm{E}+\bm{JF})\bm{s}^s_i \phantom{]}\\
    -\bm{J}\left[(\bm{C}-\bm{JD})\bm{a}^c_i + (\bm{E}+\bm{JF})\bm{s}^s_i \right]
    \end{pmatrix}\\ 
    +&
    \begin{pmatrix}
    \phantom{-\bm{J}[}(\bm{A}-\bm{JB})\bm{a}^s_i + (\bm{E}-\bm{JF})\bm{s}^c_i \phantom{]}\\
    -\bm{J}\left[(\bm{A}-\bm{JB})\bm{a}^s_i + (\bm{E}-\bm{JF})\bm{s}^c_i \right]\\
     \phantom{-\bm{J}[}(\bm{C}+\bm{JD})\bm{s}^c_i + (\bm{E}-\bm{JF})\bm{a}^s_i \phantom{]}\\
    \phantom{-}\bm{J}\left[(\bm{C}+\bm{JD})\bm{s}^c_i + (\bm{E}-\bm{JF})\bm{a}^s_i \right]
    \end{pmatrix}\\ 
    = &\lambda_i \left[
    \begin{pmatrix}
        \bm{s}^s_i\\
        \phantom{-}\bm{Js}^s_i\\
        \bm{a}^c_i\\
        -\bm{Ja}^c_i\\        
    \end{pmatrix}
    +
     \begin{pmatrix}
        \bm{a}^s_i\\
        -\bm{Ja}^s_i\\
        \bm{s}^c_i\\
       \phantom{-} \bm{Js}^c_i\\
    \end{pmatrix}
    \right].
\end{split}
\end{equation}
Here, we see that the combination of symmetric in spin and anti-symmetric in charge, or vice versa, keeps the symmetry when multiplied with $\bchi$. Since those two realizations are intrinsically orthogonal to each other, both vectors (if non-zero) must be separate eigenvectors of $\bchi$. Hence, every eigenvector of an orthogonal basis is either
\begin{equation}
   \mathbf{v}_i= \begin{pmatrix}
        \bm{s}^s_i\\
        \phantom{-}\bm{Js}^s_i\\
        \bm{a}^c_i\\
        -\bm{Ja}^c_i\\        
    \end{pmatrix} 
    \quad \text{or} \quad
    \mathbf{v}_i= \begin{pmatrix}
        \bm{a}^s_i\\
        -\bm{Ja}^s_i\\
        \bm{s}^c_i\\
       \phantom{-} \bm{Js}^c_i\\
    \end{pmatrix}.
\end{equation}

\section{Physical Basis of the BSE in the Longitudinal Channel}
\label{app:bse}
Here we briefly review the necessary transformations to obtain the BSE in \cref{eq:BSE} from the BSE in the microscopical basis of the atomic and spin indices~\cite{delre2021su2}. In the longitudinal channel the BSE reads
\begin{align}
    \label{eq:long_bse}
    \bchi^{\alpha\alpha'}_{\vq \sigma \sigma'} = &
    \left[
    \begin{pmatrix}
        \Gamma^{AA}_{\up \up} & \Gamma^{AA}_{\up \down} &  \Gamma^{AB}_{\up \up} & \Gamma^{AB}_{\up \down} \\
        \Gamma^{AA}_{\down \up} & \Gamma^{AA}_{\down \down} &
        \Gamma^{AB}_{\down \up} & \Gamma^{AB}_{\down \down} \\
         \Gamma^{BA}_{\up \up} & \Gamma^{BA}_{\up \down} &  \Gamma^{BB}_{\up \up} & \Gamma^{BB}_{\up \down} \\
        \Gamma^{BA}_{\down \up} & \Gamma^{BA}_{\down \down} &
        \Gamma^{BB}_{\down \up} & \Gamma^{BB}_{\down \down} \\
    \end{pmatrix} \right.\\ \nonumber
    &\phantom{\Biggl[}-\left.\begin{pmatrix}
    \bchi^{AA}_{0\vq\up\up} & \bm{0}&
    \bchi^{AB}_{0\vq\up\up} & \bm{0}\\
    \bm{0} & \bchi^{AA}_{0\vq\down\down} &
    \bm{0} & \bchi^{AB}_{0\vq\down\down}\\
    \bchi^{BA}_{0\vq\up\up} & \bm{0}&
    \bchi^{BB}_{0\vq\up\up} & \bm{0}\\
    \bm{0} & \bchi^{BA}_{0\vq\down\down} &
    \bm{0} & \bchi^{BB}_{0\vq\down\down}\\
    \end{pmatrix}^{-1}
    \right]^{-1}.
\end{align}
By using the symmetry $(A,\sigma) \leftrightarrow (B,-\sigma)$ the matrices in the BSE show a centro-symmetric  property in atomic and spin indices of the following form
\begin{equation}
    \bm{\mathcal{C}} =\begin{pmatrix}
        \bm{A} & \bm{B}\\
        \bm{J}\bm{B}\bm{J} & \bm{J}\bm{A}\bm{J}
    \end{pmatrix},
\end{equation}
where $\bm{A}$ and $\bm{B}$ are sub-matrices and 
\begin{equation}
    \bm{J} = \begin{pmatrix}
        \bm{0} & \mathbbm{1}\\
        \mathbbm{1} & \bm{0}
    \end{pmatrix}.
\end{equation}

$\bm{\mathcal{C}}$ can be block-diagonalized via
\begin{equation}
    \bm{\mathcal{Q}} \bm{\mathcal{C}}\bm{\mathcal{Q}}^T = 
    \begin{pmatrix}
        \bm{A}-\bm{B}\bm{J} & \bm{0}\\
        \bm{0} & \bm{A}+\bm{B}\bm{J}
    \end{pmatrix}
\end{equation}
with
\begin{equation}
    \bm{\mathcal{Q}} = \frac{1}{\sqrt{2}}
    \begin{pmatrix}
        \mathbbm{1} & - \bm{J}\\
        \mathbbm{1} & \bm{J}
    \end{pmatrix}.
\end{equation}
This gives us two independent BSE equations
\begin{align} 
    \left(\bm{\mathcal{Q}} \bchi^{\alpha\alpha'}_{\vq \sigma \sigma'}\bm{\mathcal{Q}}^T \right)_\pm \coloneqq & \bm{A}\pm\bm{B}\bm{J}\\ \nonumber
   = &\left[
   \begin{pmatrix}
       \Gamma^{AA}_{\up \up} \pm \Gamma^{AB}_{\up \down}& \Gamma^{AA}_{\up \down} \pm \Gamma^{AB}_{\up \up} \\
        \Gamma^{AA}_{\down \up} \pm \Gamma^{AB}_{\down \down} & \Gamma^{AA}_{\down \down} \pm \Gamma^{AB}_{\down \up}\\
   \end{pmatrix} \right.\\ \nonumber
  & \phantom{\Biggl[}\left.-
   \begin{pmatrix}
        \bchi^{AA}_{0\vq\up\up} & \pm \bchi^{AB}_{0\vq\up\up}\\
        \pm \bchi^{AB}_{0\vq\down\down} & \bchi^{AA}_{0\vq\down\down} 
   \end{pmatrix}^{-1}
   \right]^{-1}. 
\end{align}
Applying the similarity transformation $\bm{U} (\bm{\mathcal{Q}} \bchi^{\alpha\alpha'}_{\vq \sigma \sigma'}\bm{\mathcal{Q}}^T )_\pm \bm{U}^T$ with
\begin{equation}
    \bm{U} = \frac{1}{\sqrt{2}}
    \begin{pmatrix}
        \mathbbm{1} & -\mathbbm{1}\\
        \mathbbm{1} & \phantom{-}\mathbbm{1}
    \end{pmatrix}
\end{equation}
leads then to the physical basis of \cref{eq:BSE}.

\section{BSE in the Bethe Lattice}
\label{app:proof_bethe}
In the Bethe lattice with infinite lattice connectivity (where DMFT is exact), the BSE equation develops a simple and intuitive form due to the semi-circular density of states $\rho(\epsilon)=(1/2\pi t^2)\sqrt{4t^2-\epsilon^2}$. 
In the following, we will first re-derive the corresponding expression for the PM $(\zeta_{\nu\up}=\zeta_{\nu\down})$ phase (see also Refs.~\cite{georges1996, delre2018, reitner2020, delre2021fluctuations,kowalski2024}) and extend them to AF phase later on.
 
\subsection{PM phase}
\label{bethe_para}
For the subsequent derivation it is convenient to express the local Green's function $G_\nu$ of the Bethe lattice as Hilbert transform: 
\begin{equation}
\label{eq:bethe_g}
    G_{\nu\sigma} = \int^\infty_{-\infty}\! d \epsilon\, \rho(\epsilon) \frac{1}{\zeta_{\nu\sigma}-\epsilon} = \frac{1}{\zeta_{\nu\sigma} - t^2 G_{\nu\sigma}}
\end{equation}
with the hybridization function $\Delta_{\nu\sigma} = t^2 G_{\nu\sigma}$ and $\zeta_{\nu\sigma} = \ii\nu + \mu -\Sigma_{\nu\sigma}$. The local ``bubble" susceptibility  $\chi_0$ reads:
\begin{equation}
    \bchi_{0} = - \beta G_{\nu\sigma}^2 \delta_{\nu\nu'}
\end{equation}
and the uniform ($\vq=0$) ``bubble" susceptibility  $\bchi_{0 \vq=0}$ of the lattice is:
\begin{equation}
    \bchi_{0 \vq=0 } = - \beta \int^\infty_{-\infty}\! d \epsilon\, \rho(\epsilon) \frac{\delta_{\nu\nu'}}{\left(\zeta_{\nu\sigma}-\epsilon\right)^2} = \beta \frac{dG_{\nu\sigma}}{d\zeta_{\nu\sigma}} \delta_{\nu\nu'}.
\end{equation}
With the help of \cref{eq:bethe_g} we get
\begin{equation}
     \frac{dG_{\nu\sigma}}{d\zeta_{\nu\sigma}} = -\frac{G_{\nu\sigma}^2}{1-t^2G_{\nu\sigma}^2}
\end{equation}
and, hence, obtain for the BSE of the uniform generalized lattice susceptibility $\bchi_{\vq=0 c/s}$ the following form:
\begin{align}
    \bchi_{\vq=0 c/s} =& \left[\Gamma_{c/s} +  \bchi_{0 \vq=0}^{-1} \right]^{-1} \\ \nonumber 
    = &\left[\bchi^{-1}_{c/s} - \bchi_{0 }^{-1} +  \bchi_{0 \vq=0}^{-1} \right]^{-1} 
    = \left[\bchi^{-1}_{c/s} + \frac{t^2}{\beta}\mathbbm{1}\right]^{-1},
\end{align}
where $\Gamma_{c/s} = \bchi^{-1}_{c/s} - \bchi_{0 }^{-1}$ is the local two-particle irreducible vertex and $\bchi_{c/s}$ the local generalized susceptibility in charge or spin.

\subsection{AF phase}
In the AF phase, the Green's function becomes a matrix of the two sub-lattice sites $\alpha=(A, B)$:
\begin{align}  
\label{eq:g_af}
G_{\nu\sigma} =& \int^\infty_{-\infty}\! d \epsilon\, \rho(\epsilon) \frac{1}{\zeta_{A\nu\sigma}\zeta_{B\nu\sigma} - \epsilon^2} 
\begin{pmatrix}
    \zeta_{B\nu\sigma} & \epsilon\\
    \epsilon & \zeta_{A\nu\sigma}
\end{pmatrix}\\ \nonumber
=& \begin{pmatrix}
    G_{A\nu\sigma} & 0 \\
    0 &  G_{B\nu\sigma}
    \end{pmatrix}
    = \begin{pmatrix}
    \frac{1}{\zeta_{A\nu\sigma} - t^2 G_{B\nu\sigma}} & 0 \\
    0 &  \frac{1}{\zeta_{B\nu\sigma} - t^2 G_{A\nu\sigma}}
\end{pmatrix}
\end{align}
and the local “bubble” susceptibility $\bchi_0$ reads:
\begin{equation}
    \bchi^{\alpha\alpha}_{0 \sigma \sigma} = -\beta  G^2_{\alpha\nu\sigma} \delta_{\nu \nu'}.
\end{equation}
The respective $\bchi_{0\vq=0}$ susceptibilities become:
\begin{align}
    \bchi^{\alpha\alpha}_{0\vq=0 \sigma \sigma} =& -\beta \int^\infty_{-\infty}\! d \epsilon\, \rho(\epsilon) \frac{\zeta^2_{\overline{\alpha}\nu\sigma} \delta_{\nu \nu'}}{(\zeta_{\alpha\nu\sigma}\zeta_{\baralpha\nu\sigma}-\epsilon^2)^2}\\
    \bchi^{AB}_{0\vq=0 \sigma \sigma} =& -\beta \int^\infty_{-\infty}\! d \epsilon\, \rho(\epsilon) \frac{\epsilon^2 \delta_{\nu \nu'}}{(\zeta_{A\nu\sigma}\zeta_{B\nu\sigma}-\epsilon^2)^2},
\end{align}
where for $\alpha=A$, $\baralpha=B$ or vice versa. With the definition of
\begin{align}
    g_\nu(\lambda) &= \int^\infty_{-\infty}\! d \epsilon\, \rho(\epsilon) \frac{1}{\zeta_{A\nu\sigma}\zeta_{B\nu\sigma}-(\lambda \epsilon)^2}\\ \nonumber
    &=\frac{1}{\zeta_{A\nu\sigma}\zeta_{B\nu\sigma}(1-(\lambda t)^2 g_\nu(\lambda))}=\frac{G_{\alpha\nu\sigma}}{\zeta  _{\baralpha\nu\sigma}},\\ 
    g_\nu&\coloneqq g_\nu(\lambda=1),
\end{align}
we can re-express the susceptibilities with
\begin{align}
\label{eq:bethe_x0aa}
    \bchi^{\alpha\alpha}_{0\vq=0\sigma\sigma} =& \beta \zeta_{\baralpha\nu\sigma} \frac{d g_\nu}{d \zeta_{\alpha \nu\sigma}} \delta_{\nu\nu'}=-\beta\frac{G_{\alpha\nu\sigma}^2}{1-\varrho_\nu^2}\delta_{\nu\nu'},\\
    \label{eq:bethe_x0ab}
     \bchi^{AB}_{0\vq=0\sigma\sigma} =& -\frac{\beta}{2}\left.  \frac{d g_\nu(\lambda)}{d \lambda} \right|_{\lambda=1}\!\delta_{\nu\nu'} = -\beta \frac{\varrho_\nu g_\nu}{1-\varrho_\nu}\delta_{\nu\nu'}\\\nonumber
     =&-\beta \frac{\varrho^2_\nu }{t^2(1-\varrho_\nu^2)} \delta_{\nu\nu'},
\end{align}
where $\varrho_\nu=t^2G_{A\nu\sigma}G_{B\nu\sigma}$ and we used $1+\varrho_\nu\!=\!g_\nu \zeta_{A\nu\sigma}\zeta_{B\nu\sigma}$.
By considering the AF symmetry $(A,\sigma) \leftrightarrow (B,-\sigma)$ we can greatly simplify the $\bchi_{0}$, $\bchi_{0\vq=0}$ terms in the DMFT BSE with the help of \cref{eq:bethe_x0aa,eq:bethe_x0ab} and get:
\begin{align}
    \bchi_{\vq=0\pm} =& \left[
    \Gamma + \bchi^{-1}_{0\vq=0\pm}
    \right]^{-1} \\\nonumber
    =& \left[
    \begin{pmatrix}
        \bchi_s & \bchi_{sc}\\
        \bchi_{cs} & \bchi_{c}
    \end{pmatrix}^{-1}
    -
    \begin{pmatrix}
        \bchi_0 & \overline{\bchi}_0\\
        \overline{\bchi}_0 & \bchi_0
    \end{pmatrix}^{-1}
    \right.\\\nonumber
    &\phantom{\Biggl[}\left.
    +\begin{pmatrix}
        \bchi^{AA}_{0\vq=0}\pm \bchi^{AB}_{0\vq=0}& \overline{\bchi}^{AA}_{0\vq=0}\mp \overline{\bchi}^{AB}_{0\vq=0}\\\overline{\bchi}^{AA}_{0\vq=0}
        \pm \overline{\bchi}^{AB}_{0\vq=0} & \bchi^{AA}_{0\vq=0}\mp \bchi^{AB}_{0\vq=0}
    \end{pmatrix}^{-1}
    \right]^{-1}\\\nonumber
    =& \left[ 
    \begin{pmatrix}
        \bchi_s & \bchi_{sc}\\
        \bchi_{cs} & \bchi_{c}
    \end{pmatrix}^{-1}  
    \pm\frac{t^2}{\beta} 
    \begin{pmatrix}
         \mathbbm{1} & 0\\
        0 & - \mathbbm{1}
    \end{pmatrix}
    \right]^{-1}\!.
\end{align}

\section{Bethe Lattice: Interaction Expansion of the Local Susceptibility in a Magnetic Field}
\label{app:bethe_expansion}
To get an intuitive understanding of the different effects that the magnetic field has on the local generalized susceptibility in weak and strong coupling, we consider an expansion of the local generalized susceptibility in the Bethe lattice in terms of the interaction $U$.

\subsection{Lowest order ($U=0$)}
In the non-interacting $U=0$ case, or in the lowest order, the local generalized susceptibility $\bchi$ of the Bethe lattice is given by the bare local ``bubble" $\bchi^{\alpha\alpha(0)}_{0\sigma \sigma} = -\beta (G^0_{\alpha\nu\sigma})^2\delta_{\nu\nu'}$. For a magnetic field $h$, the local susceptibility in the charge and spin components then reads
\begin{equation}
    \bchi^{(0)}_0 = \begin{pmatrix}
        \bchi^{\alpha\alpha(0)}_{0} & \overline{\bchi}^{\alpha\alpha(0)}_{0}\\
        \overline{\bchi}^{\alpha\alpha(0)}_{0} & \bchi^{\alpha\alpha(0)}_{0}
    \end{pmatrix}.
\end{equation}
When $h$ is staggered (antiferromagnetic with $h_{A} = -h_{B}$), the Green's function in the Bethe lattice becomes
\begin{equation}
    G^0_{\alpha \nu \sigma} = g^0_\nu \zeta_{\overline{\alpha}\nu\sigma} = g^0_\nu (\ii \nu + \mu -\operatorname{sgn}(\sigma)h_\alpha), 
\end{equation}
with 
\begin{align}
    g^0_\nu &= \frac{1}{(\ii \nu + \mu +h)(\ii \nu + \mu -h) (1-t^2 g^0_\nu)} \\\nonumber
    &= \frac{1}{2t^2} \varmp \sqrt{\frac{1}{4t^4}-\frac{1}{t^2 (\ii \nu + \mu +h)(\ii \nu + \mu -h)}},
\end{align}
where the $\varmp$ has to be chosen such that the Green's function obeys the correct asymptotics for $\nu \to \pm \infty$.
The ``bubble" reads 
\begin{align}
\label{eq:bare_bubble}
    \bchi \stackeq{U=0}& \bchi^{(0)}_0 \\ \nonumber
    = & \beta \delta_{\nu\nu'} (g^0_\nu)^2 \begin{pmatrix}
        -(\ii\nu+ \mu)^2-h^2 & 2 h (\ii\nu + \mu)\\
        2 h (\ii\nu + \mu) &  -(\ii\nu+ \mu)^2-h^2
    \end{pmatrix}.
\end{align}
For half-filling ($\mu=0$) the eigenvalues of $\bchi^{(0)}_0$ are 
\begin{equation}
    \lambda_\pm = \beta (g^0_\nu)^2 (\nu^2-h^2 \pm 2 \ii \nu h)
\end{equation}
displaying a ``fountain"-like structure when plotting their real and imaginary part for finite magnetic field $h$, similar to panel (c) of \cref{fig:val_compare}. Here, \cref{eq:bare_bubble} clearly shows the suppressive effect $ \propto (\nu^2-h^2)$ of the magnetic field $h$ on the charge and spin susceptibility.

\subsection{First order in $U$}

\begin{figure}
\begin{align*}
  \bchi_{c} =  \frac{1}{2}\sum_\sigma \Bigg(-
\begin{tikzpicture}[baseline=0.cm]
        \begin{feynhand}
            \node at (-0.7,-0.3) {$\sigma$};
            \node at (-0.7,0.3) {$\sigma$};
            \node at (0.7,-0.3) {$\sigma$};
            \node at (0.7,0.3) {$\sigma$};
            \vertex (O) at (0,0);
            \vertex (a) at (-0.5,0.3); 
            \vertex (c) at (0.5,0.3);
            \vertex (d) at (-0.5,-0.3);
            \vertex (e) at (0.5,-0.3);
            \propag [fer] (a) to (c);
            \propag [fer] (e) to (d);
        \end{feynhand}
\end{tikzpicture}
 & -
\begin{tikzpicture}[baseline=0.cm]
        \begin{feynhand}
            \node at (-0.7,-0.3) {$\sigma$};
            \node at (-0.7,0.3) {$\sigma$};
            \node at (0.7,-0.3) {$\sigma$};
            \node at (0.7,0.3) {$\sigma$};
            \vertex (O) at (0,0);
            \vertex (a) at (-0.5,0.3); 
            \draw [with arrow=0.25] (0,0.5) circle (0.2);
            \vertex [gray,dot] (b) at (0,0.3){};
            \vertex (c) at (0.5,0.3);
            \vertex (d) at (-0.5,-0.3);
            \vertex (e) at (0.5,-0.3);
            \propag [fer] (e) to (d);
            \propag [fer] (a) to (b);
            \propag [fer] (b) to (c);
        \end{feynhand}
\end{tikzpicture}\\
-
\begin{tikzpicture}[baseline=0.cm]
        \begin{feynhand}
            \node at (-0.7,-0.3) {$\sigma$};
            \node at (-0.7,0.3) {$\sigma$};
            \node at (0.7,-0.3) {$\sigma$};
            \node at (0.7,0.3) {$\sigma$};
            \vertex (O) at (0,0);
            \vertex (a) at (-0.5,0.3); 
            \vertex (c) at (0.5,0.3);
            \vertex (d) at (-0.5,-0.3);
            \vertex (e) at (0.5,-0.3);
            \draw [with arrow=0.25] (0,-0.1) circle (0.2);
            \vertex [gray,dot] (b) at (0,-0.3){};
            \propag [fer] (e) to (b);
            \propag [fer] (b) to (d);
            \propag [fer] (a) to (c);
        \end{feynhand}
\end{tikzpicture}
&-
\begin{tikzpicture}[baseline=0.cm]
        \begin{feynhand}
            \node at (-0.7,-0.3) {$\sigma$};
            \node at (-0.7,0.3) {$\sigma$};
            \node at (0.8,-0.3) {$-\sigma$};
            \node at (0.8,0.3) {$-\sigma$};
            \vertex (O) at (0,0);
            \vertex (a) at (-0.5,0.3); 
            \vertex (c) at (0.5,0.3);
            \vertex (d) at (-0.5,-0.3);
            \vertex (e) at (0.5,-0.3);
            \vertex [gray,dot] (O) at (0,0){};
            \propag [fer] (a) to (O);
            \propag [fer] (O) to (c);
            \propag [fer] (e) to (O);
            \propag [fer] (O) to (d);
        \end{feynhand}
\end{tikzpicture}
+ \dots \Bigg)
\end{align*}
\caption{Diagrammatic representation of the interaction expanded local charge susceptibility, where $G^0_{\nu\sigma}=\fermion$ and $U=\vertex$.
}
    \label{fig:expansion}
\end{figure}

The first-order corrections in $U$ to the local generalized charge susceptibility are displayed in \cref{fig:expansion}, reading
\begin{align}
    \bchi^{(1)}_{0\sigma\sigma} =& -2U \beta \delta_{\nu\nu'}  (G^0_{\nu\sigma})^3 \left(\frac{1}{\beta} \sum_{\nu_1} \e^{\ii\nu0^+} G^0_{\nu_1-\sigma}\right),\label{eq:1st_hartree}\\
    \bchi^{(1)}_{\sigma-\sigma} =& - U (G^0_{\nu\sigma})^2 (G^0_{\nu'-\sigma})^2. \label{eq:1st_vertex}
\end{align}
The former expression in \cref{eq:1st_vertex} corresponds to the first self-energy correction of $\bchi_0$ and the latter in \cref{eq:1st_vertex}  to the first vertex correction, namely the bare vertex. At half-filling and zero magnetic field the first self-energy correction is compensated by the chemical potential $\mu=U/2$ and the vertex correction yields a negative contribution, which for larger values of $U$ can yield negative eigenvalues in the local charge susceptibility 
\begin{equation}
    \bchi_c^{(1)} = \frac{1}{2}\sum_{\sigma}\left( \bchi^{(0)}_{0\sigma\sigma}+\bchi^{(1)}_{0\sigma\sigma}+\bchi^{(1)}_{\sigma-\sigma}\right)
\end{equation}
(although in lowest order the physical charge response becomes then negative as well). 

It is then instructive to consider the change of the different contributions to $\chi_c^{(1)} = (1/\beta^2)\sum_{\nu\nu'}\bchi_c^{(1)}$ when turning on a staggered magnetic field $h$: (i) $\chi^{(0)}_{0}=(1/2\beta^2)\sum_{\nu\nu'\sigma}\bchi^{(0)}_{0\sigma\sigma}$ remains positive and becomes suppressed, as discussed in the preceding section, (ii) $\chi^{(1)}_{0}=(1/2\beta^2)\sum_{\nu\nu'\sigma}\bchi^{(1)}_{0\sigma\sigma}$ will become non-zero and negative, further suppressing $\chi^{(1)}_c$, and (iii) $(1/2)\sum_\sigma \chi^{(1)}_{\sigma-\sigma}=(1/2\beta^2)\sum_{\nu\nu'\sigma}\bchi^{(1)}_{\sigma-\sigma}$ becomes \emph{less} negative, since $(G^0_{\nu\sigma})^2$ is getting smaller. The overall change of the charge response in first-order with increasing $U$ is thus depending on the delicate balance between the suppressive term of $\chi^{(1)}_0$ and the effectively \emph{enhancing} term of the vertex in $\chi^{(1)}_c$. For small values of the interaction, the suppressive effect of $\chi^{(0)}_{0}$ will generally dominate the behavior of $\chi_c$. With increasing $U$ at $h=0$, the vertex contributions dominate and lead to negative eigenvalues in the charge susceptibility also for higher order in $U$. For finite magnetic field $h>0$, the balance between the suppressive and enhancing contributions depending on the lattice and magnetic field in $G^0_{\nu\sigma}$ then determines the overall behavior of the local charge response at strong coupling. The delicate balance can also be made responsible for the overall suppressive effect of $h$ in the RPA approximation; there, the self-energy corrections are not treated on equal footing with the vertex corrections (not in the same order of $U$), not allowing for negative eigenvalues and tipping the balance to the suppressive side. 

On the other hand, in DMFT, the correct treatment of interaction diagrams order per order in the Monte Carlo sampling of the auxiliary AIM can tip the balance to the vertex corrections, and the magnetic field can lead to an enhancement of the local charge response in the regime of the Heisenberg antiferromagnet.

\section{Charge response in the atomic limit at \ph-symmetry}
\label{app:al_charge_resp}
For the atomic limit of the Hubbard model (AL) in a magnetic field $h$ at \ph-symmetry $\mu=U/2$
\begin{equation}
    H = - \frac{U}{2} \left(\hat{n}_{\up}+ \hat{n}_{\down}\right) -h \left(\hat{n}_{\up}- \hat{n}_{\down}\right) +U  \hat{n}_{\up} \hat{n}_{\down},
\end{equation} the local charge response $\chi_c$ becomes always suppressed for $\abs{h}>0$, in comparison with $h=0$:
\begin{equation}
    \chi_c = \frac{\beta }{e^{\frac{\beta  U}{2}} \cosh (\beta  h)+1},
\end{equation}
since $\cosh (\beta  h)\!>\!1$ for $\abs{h}\!>\!0$.
This is in contrast to the strong coupling limit of the Heisenberg AF in DMFT, where $\chi_c$ becomes enhanced in the AF phase (as discussed in \cref{sec:loc_resp}). Although, the eigenspectra of the local generalized susceptibilities $\bchi$ look qualitatively similar between the AL and the Heisenberg AF: Both show similar eigenvalues---including negative ones---and have similar eigenvectors. This difference in the charge response $\chi_c$ illustrates that negative eigenvalues are a necessary requirement---since they stem from the dominant negative vertex contributions---allowing for the local enhancement of the charge response in a magnetic field. However, they alone may not be sufficient. The behavior of the physical response, suppressive or repulsive, still depends on the overall balance between the bubble $\bchi_0$ and the vertex contributions to $\chi_c$.

\section{Local spin response}
\label{app:spin_response}
\begin{figure}[th]
    \centering
    \includegraphics[width=\textwidth]{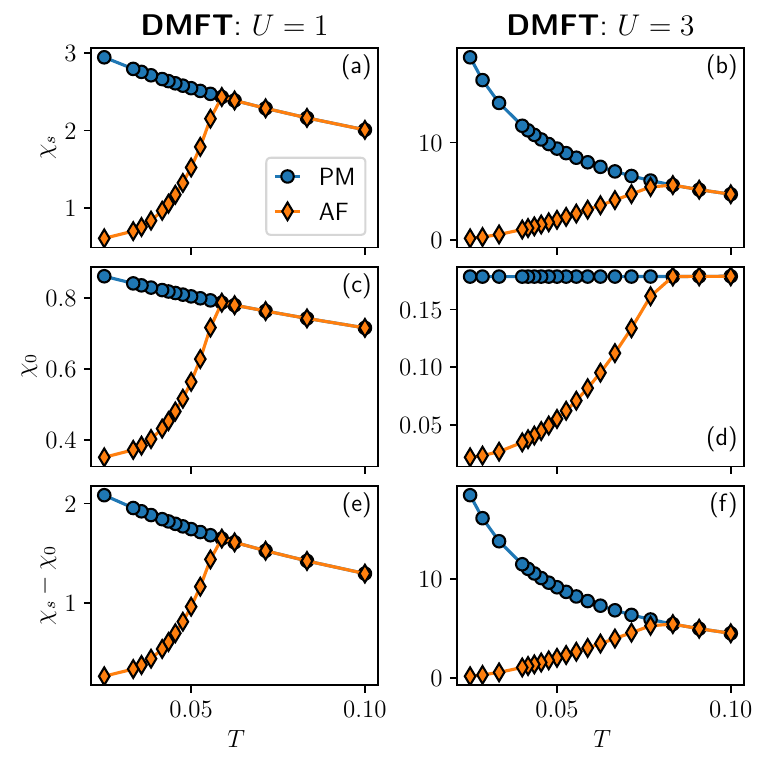}
    \caption{Comparision of the local spin response $\chi_s$ and of its different diagrammatic contributions as a function of temperature $T$ in our anti-ferromagnetic (AF, orange diamonds) and paramagnetic (PM, blue circles) DMFT calculations for weak coupling $U=1$ (left column) and strong coupling $U=3$. Top row: local spin response $\chi_s$. Middle row: bubble contribution $\chi_0$ to $\chi_s$. Bottom row: the difference between the bubble contribution $\chi_0$ and $\chi_s$, corresponding to the vertex corrections.}
    \label{fig:loc_spin}
\end{figure}

In contrast to the local charge response $\chi_c$ (see \cref{sec:loc_resp}) the  local spin response  $\chi_s=\frac{1}{\beta^2}\sum_{\nu\nu'}\bchi_s$, calculated in DMFT, decreases in the weak coupling Slater AF and in the strong coupling Heisenberg AF regime w.r.t.~the PM solution. The qualitative similar effect of the AF order onto $\chi_s(T)$ for two representative values for weak and strong coupling, $U=1$ and $U=3$ respectively, can be observed in the top row of \cref{fig:loc_spin}, where we compare $\chi_s$ , similar to $\chi_c$ in \cref{fig:local_ch_rsp}. Since the bubble contribution $\chi_0$ to $\chi_s$ (middle row) is the same as for $\chi_c$, the difference in the behavior stems from the vertex contributions $\chi_{\text{vert}}=\chi_s-\chi_0$ (bottom row). In contrast to the case of $\chi_c$, for $\chi_s$ the vertex corrections $\chi_{\text{vert}}$ have a positive value and decrease in the AF phase w.r.t.~the PM solution, showing a similar suppressive behavior as $\chi_0$.

This different behavior can be already understood at the level of the first-order vertex correction $\bchi^{(1)}_{\sigma-\sigma}$ (see \cref{app:bethe_expansion}), which enters $\chi_s$ with the opposite sign in comparison to $\chi_c$ (cf. \cref{eq:loc_chi_s}). At the same time, additional negative real eigenvalues of $\bchi$ in the AF Heisenberg regime, stemming from spin dominated vertex divergence lines, have an additional suppressive effect on $\chi_s$. Thus, they further contribute to the suppression of $\chi_s$ in the AF strong coupling regime.

\FloatBarrier
\bibliography{Ref} 

\end{document}